 \documentclass[preprint,review,12pt]{elsarticle}

 \usepackage{multirow}
 \usepackage{tabularx}
 \usepackage{rotating}
 \usepackage{setspace}

 \usepackage{graphicx}
 \usepackage{epsfig,subfigure,psfrag}

\usepackage{textcomp}
\usepackage{latexsym,amssymb,amsmath}
 \usepackage{lineno}

\begin{document}

\begin{frontmatter}



\title{Improving the staggered grid Lagrangian hydrodynamics
for modeling multi-material flows}


 \author[a]{Hai-bo Zhao}
 \author[a]{Bo Xiao\corref{cor1}}\ead{homenature@pku.edu.cn}
 \author[a]{Jing-song Bai\corref{cor1}}\ead{bjsong@foxmail.com}
 \author[a]{Shu-chao Duan}
 \author[a]{Gang-hua Wang}
 \author[a]{Ming-xian Kan}

\cortext[cor1]{Corresponding author}
\address[a]{Institute of Fluid Physics, CAEP, P. O. Box 919-105, Mianyang 621900, China}

\begin{abstract}
In this work, we make two improvements on the staggered grid hydrodynamics (SGH) Lagrangian scheme for modeling 2-dimensional compressible multi-material flows on triangular mesh. The first improvement is the construction of a dynamic local remeshing scheme for preventing mesh distortion. The remeshing scheme is similar to many published algorithms except that it introduces some special operations for treating grids around multi-material interfaces.
This makes the simulation of extremely deforming and topology-variable multi-material processes possible, such as the complete process of a heavy fluid dipping into a light fluid. The second improvement is the construction of an Euler-like flow on each edge of the mesh to count for the ``edge-bending'' effect, so as to mitigate the ``checkerboard'' oscillation that commonly exists in Lagrangian simulations, especially the triangular mesh based simulations. Several typical hydrodynamic problems are simulated by the improved staggered grid Lagrangian hydrodynamic method to test its performance.
\end{abstract}


\end{frontmatter}


%%\section{}
%%\label{}

\section{Introduction}\label{sec.introduction}

In the simulation of fluids, there exist two kinds of frames that are the bases of simulation methods, one is the Eulerian frame where the partial differential equations are discretized on the fixed grids, and the other is the Lagrangian frame where grids move with the fluid.
In Lagrangian approach, the mass advection term vanishes due to the consistence of the frame with the flow, and the discontinuity in multi-material fluid flows can be sharply captured, which is a superiority of the Lagrangian approach to the Eulerian approach.
The geometric conservative law is a cornerstone that should be satisfied in any Lagrangian scheme \cite{Marie:2009}.
The nature way to achieve this goal is to employ a staggered discretization in which the position, velocity and kinetic energy are centered at vertices, while density, pressure and internal energy are within cells \cite{Neumann:1950, Wilkins:1964}.
We choose the SGH Lagrangian scheme for modeling 2-dimensional multi-material flows in this work.

One major difficulty of the Lagrangian simulation is mesh distortion.
For example, when simulating the vortex, the cells will become too much distorted to be suitable for computation.
This failure is due to the combination of two ingredients \cite{Stephane:2011}: in one hand, the use of a fixed-connectivity mesh, and in the other hand, the absence of mass fluxes between cells obtained with the Lagrangian formulation.
The Arbitrary Lagrangian-Eulerian (ALE) method, which allows the freedom of an arbitrary relative movement between the mesh and the fluid, is an option to relax the distortion.
However, if distorted grids emerge around the large deforming interfaces in multi-material
flow, they usually cannot be eliminated by the normal ALE approach.
In this case, the multi-material ALE (MMALE) \cite{Anbarlooei:2009, Tian:2011, QinghongZeng:2014} or reconnection-based ALE (ReALE) \cite{Loubere:2010,Bo:2015,JieLiu:2016} are applied.

The mesh distortion can also be handled by dynamic remeshing in Lagrangian simulations.
The most attractive approach of them is the dynamic local remeshing \cite{Martin:2010}, which utilizes some basic grid operations such as ``edge-splitting'', ``edge-swapping'', and ``edge-merging'' to eliminate low quality grids.
This method is usually based on triangular mesh in 2D and tetrahedral mesh in 3D.
Compared to the global remeshing scheme \cite{Tscharner:2015} which replaces the whole mesh with a new one or the half-global remeshing scheme \cite{Lin:2011} which
cuts out a distorted region and fills it with a new mesh, the dynamic local remeshing can make least modifications to the grids and so is most efficient and brings least remeshing error.
Up to now, many researchers have proposed various kinds of dynamic local remeshing algorithms \cite{Stephane:2011,Martin:2010,Clausen:2013,Dapogny:2014,Baker:2005} that are effective
for handling mesh distortion or realizing adaptive mesh in Lagrangian simulations.
But most of those algorithms are applied for single material.
To extend those algorithms to deal with grids around multi-material interfaces is not easy:
it is not always possible to keep grids around an interface in good quality without modifying the interface, and algorithms are needed to treat new cells covering multiple materials.
In this paper, we are trying to construct a 2-dimensional dynamic local remeshing scheme that includes the treatment of interface grids for multi-material flows.

Our second improvement focuses on the ``checkerboard'' oscillation problem that commonly exists for Lagrangian simulations, especially the triangular mesh based simulations.
We trace the origin of the checkerboard oscillation back to the non-bending of the grid edges when the mesh move with the flow, and introduce an Euler-like flow for each edge to compensate for this edge-bending effect.
This compensation term is simple, but is shown to be quite effective to mitigate the checkerboard oscillation in practice.

The remaining paper is organized as follows.
First, the Lagrangian staggered grid hydrodynamics (SGH) with first-order spatial discretization is presented in section \ref{sec.SGH}.
Then, section \ref{sec.remeshing} is devoted to the discussion of a dynamic remeshing scheme that is applied for
conquering the mesh distortion problem in SGH Lagrangian simulations.
In section \ref{sec.FixOscillation}, the oscillation problem is analyzed and a compensation matter flow method is proposed to fix it.
In Section \ref{sec.NumericalTests}, we present some numerical experiments that enlighten the good behavior of the method.
Finally, we conclude and give some directions for future works.

\section{Conservation laws and SGH discretization}\label{sec.SGH}

The work of this paper is based on the SGH Lagrangian hydrodynamics.
This section gives a brief introduction to this scheme.

The conservative equations of momentum and internal energy in Lagrangian frame are shown in Eqs. (1) and (2) respectively.
\begin{eqnarray}
  \frac{d}{dt}\int_V \rho \vec{u} dV &=& \oint_{\partial V}(-p)\vec{n}dA, \\
  \frac{d}{dt}\int_V \rho e dV  &=& (-p)\oint_{\partial V}\vec{n}\cdot\vec{u} dA,
\end{eqnarray}
with $\rho$ the density, $u$ the velocity, $p$ the pressure, and $e$ the internal energy.

\begin{figure}[htpb]
  \centering
  \includegraphics[width=0.3\textwidth]{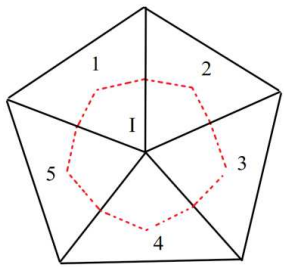}\\
  \caption{SGH discretization.}\label{fig.SGHdiscretization}
\end{figure}

In  SGH approach, the equations are discretized on different control volumes \cite{Nathaniel:2013}.
The discrete form of the conservative momentum is:
\begin{equation}
  M_I\frac{\Delta \vec{u}}{\Delta t} = \sum_{\text{MCV}}(-p)\vec{n}\delta A,
\end{equation}
where the MCV donates the momentum control volume.
The velocity is located at the vertex, so the dual grid, on which the momentum is integrated, should be constructed.
In Figure \ref{fig.SGHdiscretization}, the midpoints of the cell edges and the center of the triangle are connected by the dash line to form the momentum control volume.

\begin{figure}[htpb]
  \centering
  \includegraphics[width=0.9\textwidth]{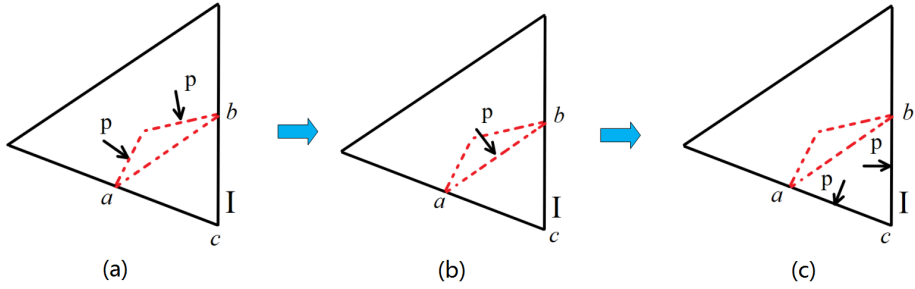}\\
  \caption{Pressure forces acting on the control volume is equivalent to forces acting on triangle edges.}\label{fig.SGHPressure}
\end{figure}

In our simulation, first order scheme is adopted for space discretization, then the cell variables (pressure, density, and internal energy) are piecewise constant. So, in Fig. \ref{fig.SGHdiscretization}, the mass of the vertex I is one third of the sum of the triangles surrounding the vertex I. That is, $M_{\rm I}=1/3*(M_1+M_2+M_3+M_4+M_5)$. Also, based on edge vectorial resultant, the pressure $p$ acting on the control volume (Fig. \ref{fig.SGHPressure}(a)) is equivalent to the pressure acting on \underline{ab} (Fig. \ref{fig.SGHPressure}(b)), which also can be decomposed to the pressure acting on \underline{ac} and \underline{bc} (Fig. \ref{fig.SGHPressure}(c)). This equivalence is convenient for programming.

The discrete form of the conservative internal energy is:
\begin{equation}
  M \frac{\Delta e}{\Delta t} = (-p)\sum_{\text{ECV}} \vec{n}\cdot\vec{u} \delta A,
\end{equation}
where the ECV donates the internal energy control volume, which is the cell.

\begin{figure}[htpb]
  \centering
  \includegraphics[width=0.3\textwidth]{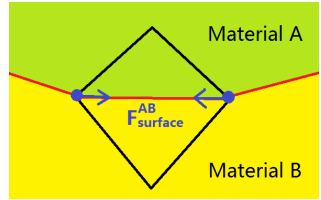}\\
  \caption{Surface tension force realization in SGH.}\label{fig.SurfaceTension}
\end{figure}

A surface tension on the multi-material interface is realized by applying a force $F^{\text{AB}}_{\text{surface}}$ on each edge on the interfaces, as illustrated by Fig. \ref{fig.SurfaceTension}. The size of $F^{\text{AB}}_{\text{surface}}$ is determined by the material types A and B on both sides of the interface.

Finally, to stabilize the discrete hydrodynamic simulations, a viscosity is required.
A viscosity tensor $\sigma_{\text{viscos}}=c \times \dot{\varepsilon} \times (\rho/\rho_0)$ is applied in our simulations,
where $c$ is the viscosity coefficient, $\dot{\varepsilon}$ the strain rate tensor, and $\rho_0$ and $\rho$ are the initial and current mass densities.

\section{2D dynamical local remeshing}\label{sec.remeshing}

A dynamical local remeshing scheme is constructed in this section for prohibiting mesh distortion in SGH Lagrangian simulations.
The remeshing algorithm is based on pure triangular mesh.

\begin{figure}[htpb]
  \centering
  \subfigure[]{\includegraphics[width=0.25\textwidth]{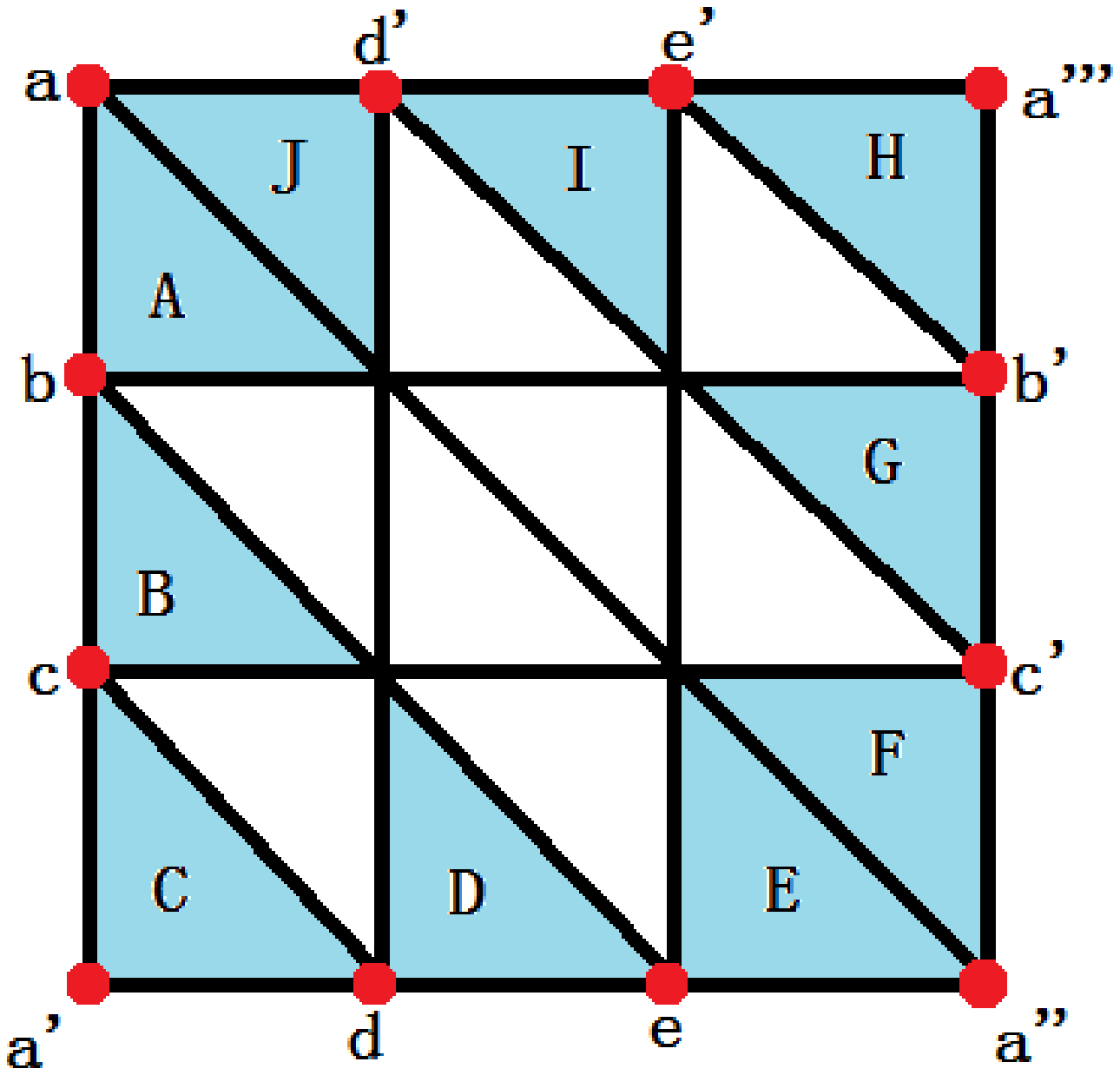}\label{fig.CycledMesh}}
  \ \ \ \
  \subfigure[]{\includegraphics[width=0.25\textwidth]{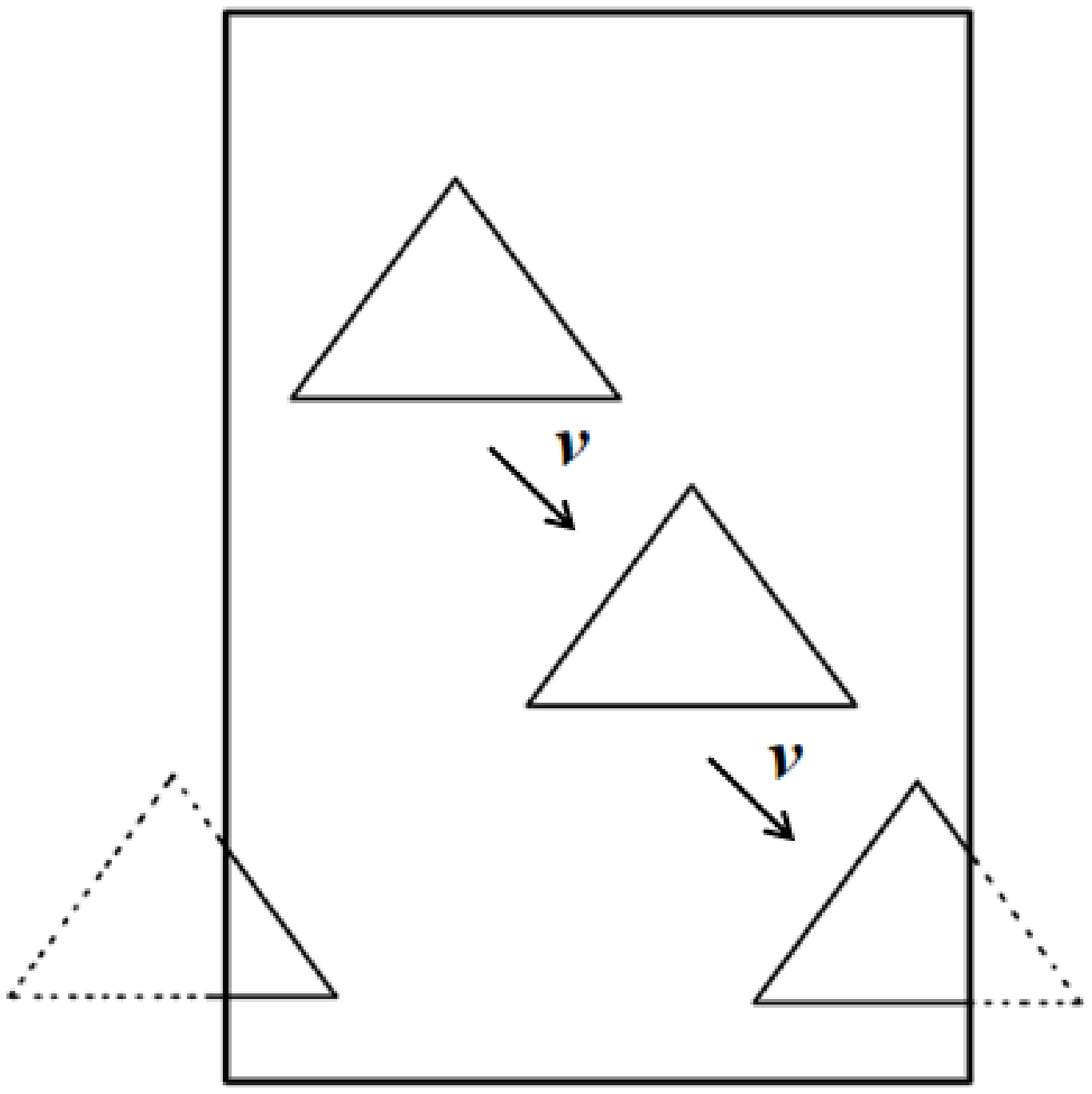}\label{fig.CycledPhysics}}\\
  \caption{(a) Illustration of the cycled mesh boundary condition; (b) Illustration of realizing physical cycled boundary condition on cycled mesh.}\label{fig.CycledMeshConditions}
\end{figure}

Before discussing the remeshing scheme, it is worth mentioning about here the ``cycled mesh boundary condition''.
For the simulations in this work, mesh always fills into a rectangle box at the initial state.
Usually, the outside neighbor of a cell on border is set null.
But for the ``cycled mesh boundary condition'', the outside neighbor of a border cell is set as
the corresponding cell on the opposite border, which equivalently means, a cell on border shares a same edge and a same pair of vertices with its corresponding cell on the opposite border.
This is illustrated in Fig. \ref{fig.CycledMesh}:
the vertices $a$, $a'$, $a''$ and $a'''$ are the same vertex, $b$ and $b'$ are the same vertex, and so on;
the triangles $A$ and $H$ are neighbours, $B$ and $G$ are neighbours, and so on.
This type of cycled mesh brings some conveniences to our work,
one is that it is natural to realize cycled physical boundary condition on the mesh, as illustrated in Fig. \ref{fig.CycledPhysics},
another is that the remeshing operations do not have to distinct between
inner and border cells, which facilitates the programming.

\subsection{The four basic grid operations}

\begin{figure}[htpb]
  \centering
  \subfigure[]{\includegraphics[width=0.45\textwidth]{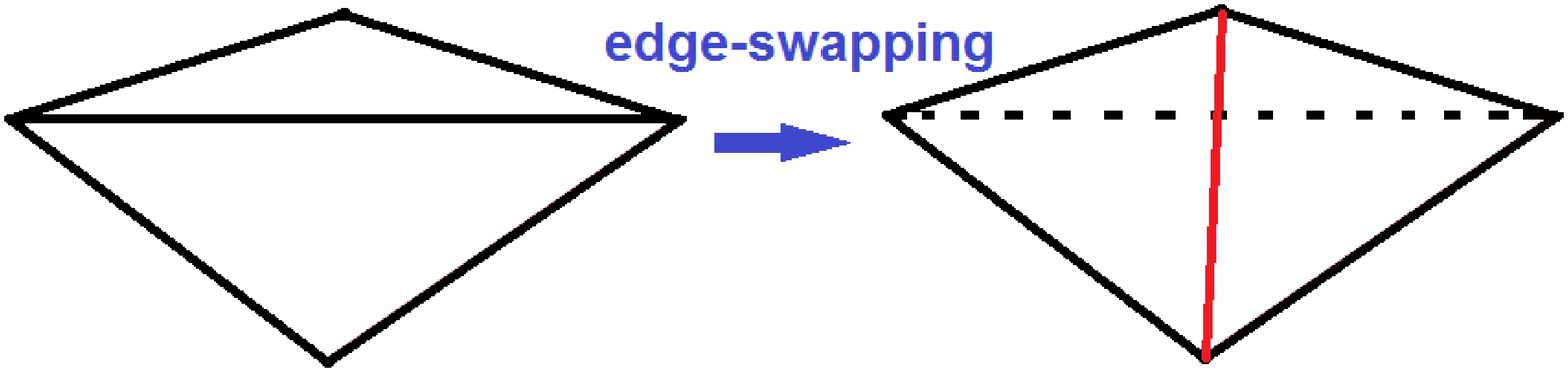}\label{fig.EdgeSwap}}
  \ \ \ \
  \subfigure[]{\includegraphics[width=0.45\textwidth]{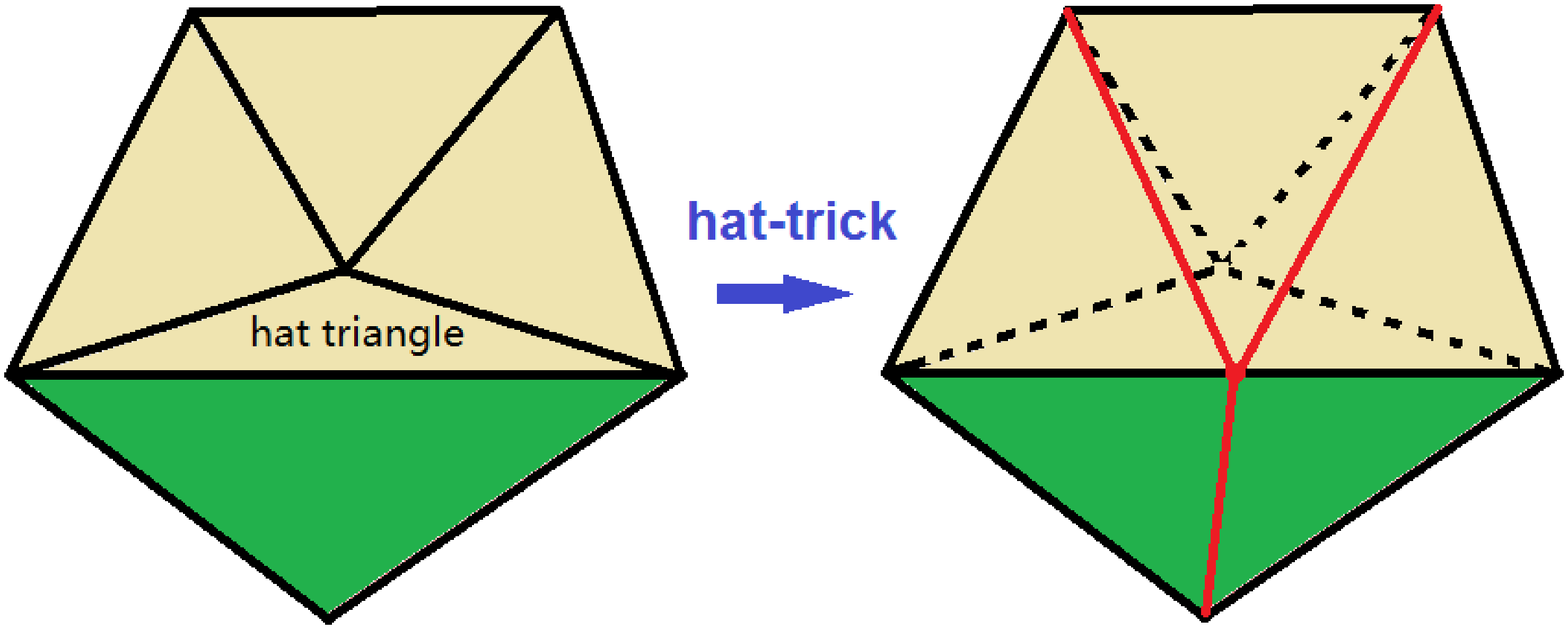}\label{fig.HatTrick}}\\
  \subfigure[]{\includegraphics[width=0.45\textwidth]{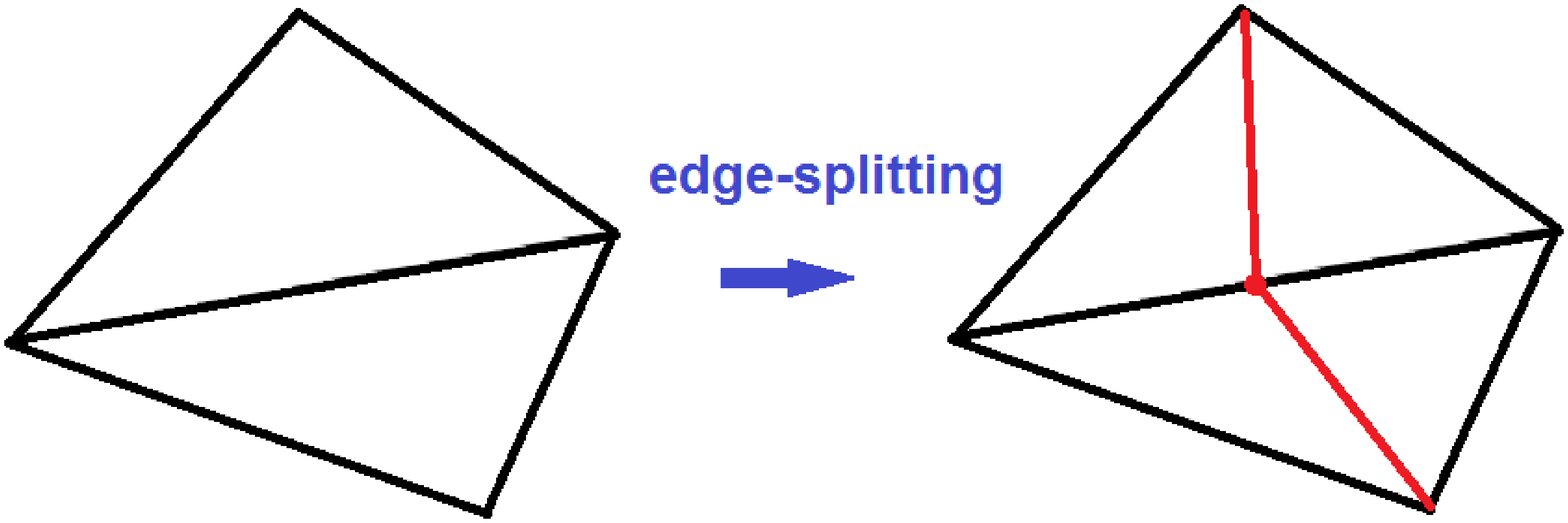}\label{fig.EdgeSplit}}
  \ \ \ \
  \subfigure[]{\includegraphics[width=0.45\textwidth]{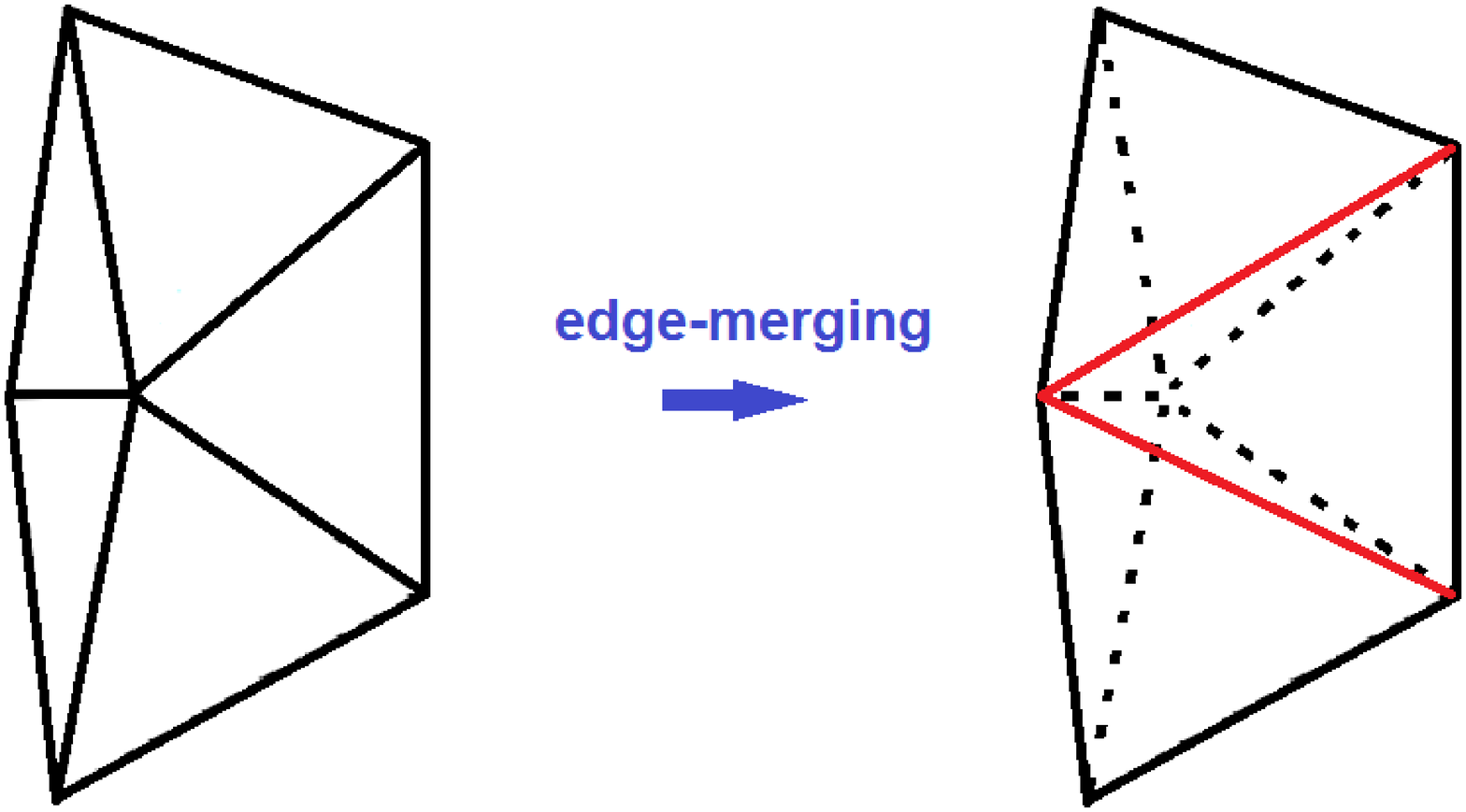}\label{fig.EdgeMerge}}\\
  \caption{The four basic grid operations.}\label{fig.FourOperations}
\end{figure}

The remeshing scheme in this work is constructed from four basic operations: edge-swapping, edge-splitting, edge-merging and ``hat-trick'', as shown in Fig. \ref{fig.FourOperations}.
The former three operations are familiar in literatures,
and they are extended to include the treatment of the grids around multi-material interfaces in this work.
The ``hat-trick'' is a new operation proposed in this work, which is used specially for grids on interfaces.

Edge swapping tries to improve the mesh quality by a local reconnection of the vertices.
It comes up when the maximum angle of a triangle is larger than some threshold,
but if the pair of triangles to be swapped possess different materials, stricter criterions need to be satisfied.
Hat-trick deletes a flat ``hat triangle'' by moving a vertex on top of the ``hat'' to the \emph{midpoint} of its bottom edge.
On the interface, hat-trick is preferred than edge-swapping, unless the hat-trick is prevented by some extra criterions.
Edge-splitting inserts a new vertex in the \emph{midpoint} of the longest edge of a triangle when the edge length is above some threshold.
The threshold for edge-splitting is larger inside a material than on an interface.
Edge-merging is opposite to edge-splitting,
and it deletes a vertex at the end of the shortest edge of a triangle when the edge is below some threshold.
The threshold for edge-merging is also larger inside a material than on an interface.
More explicit descriptions of criterions for those operations are found in the remeshing algorithm of Sec. \ref{sec.RemeshAlgorithm}.

\subsection{Remapping strategies for the operations}\label{sec.Remap}

The physical quantity remapping strategies for the operations are based on the area weighted averages.
First, the coefficients $\{c_n^k\}$ describing the overlapping area of the new triangles $\{k\}$ to old triangles $\{n\}$ are obtained.
Then the cell integration quantities, i.e. the area $S$, mass $M$ and internal energy $E$, of the new triangles $\{k\}$ are calculated by the weighted summation of those of the old triangles:
\begin{equation}\label{eq.NormalRemap}
  \begin{split}
  S_{\text{new}}^k &= \sum_n c_n^k \cdot S_{\text{old}}^n\\
  M_{\text{new}}^k &= \sum_n c_n^k \cdot M_{\text{old}}^n\\
  E_{\text{new}}^k &= \sum_n c_n^k \cdot E_{\text{old}}^n
  \end{split}
\end{equation}
For edge-splitting, the coefficients $\{c_n^k\}$ take the values of $0.5$ and $0.0$.
For edge-swapping, the intersection point of the new diagonal line and the old one needs to be calculated for deriving the coefficients $\{c_n^k\}$.
The most non-trivial task is to obtain $\{c_n^k\}$ for edge-merging or hat-trick, where complex intersection points need to be calculated.
Once the cell integration quantities are obtained, the cell strength quantities, i. e. the specific internal energy $e$, density $\rho$ and pressure $p$, are calculated by
\begin{equation}
  \begin{split}
  e^k &= E^k / M^k\\
  \rho^k &= M^k / S^k\\
  p^k &= \text{EOS}(e^k, \rho^k)
  \end{split}
\end{equation}

If the operations of edge-swapping and edge-merging take place on the interface,
where a new triangle may overlaps old triangles containing a different material,
the remapping strategies need to be modified.
There are many discussions in literatures about how to treat a cell containing mixed materials.
In this work, we try to adopt the simplest approach: a triangle is always forced to contain single material.
Based on this, the remapping strategies are designed to ``push'' the matter into the neighboring triangles that contain the same material.
In seldom cases, there is no place to push the matter into, then the matter has to be thrown away.
This kind of remapping strategies would inevitably bring some non-physical effects,
and in order to suppress those effects, the remeshing algorithm in Sec. \ref{sec.RemeshAlgorithm}
applies stricter criterions for the operations to take place on interface.
The following paragraphs explain more about the remapping strategies for edge-swapping and edge-merging on interfaces.

\begin{figure}[htpb]
  \centering
  \includegraphics[width=0.6\textwidth]{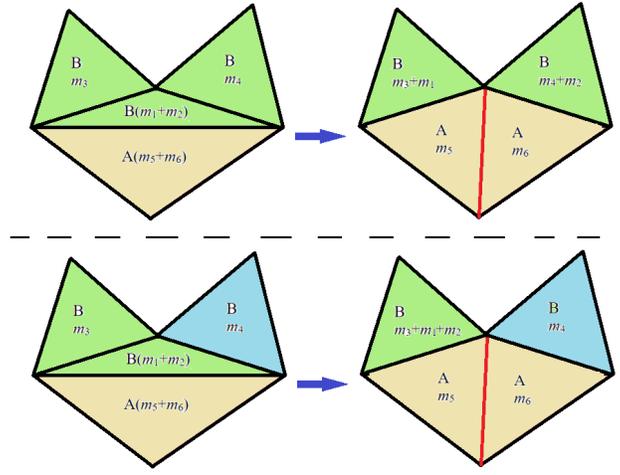}\\
  \caption{Remap for edge-swapping on interface}\label{fig.SwapMapOnInterface}
\end{figure}

Fig. \ref{fig.SwapMapOnInterface} shows two examples of edge-swapping on the interface.
The matters assignment is as follows:
the quadrilateral $\square abcd$ is filled with the material A,
and the mass of the material B is transferred to the neighboring triangles that possess the same material.

\begin{figure}[htpb]
  \centering
  \includegraphics[width=3.0in]{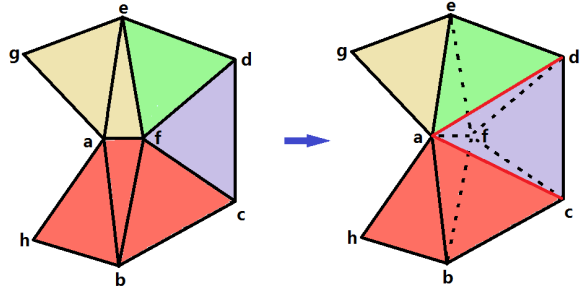}
  \caption{Remap for edge-merging on interface.}\label{fig.MergeMapOnInterface}
\end{figure}

Fig. \ref{fig.MergeMapOnInterface} shows an example of edge-merging on the interface.
The matters assignment is as follows.
For the convenience of description,
the new triangles $\bigtriangleup abc$, $\bigtriangleup acd$ and $\bigtriangleup ade$ are named the ``son'' of
the corresponding old triangles $\bigtriangleup fbc$, $\bigtriangleup fcd$ and $\bigtriangleup fde$, respectively.
The remapping scheme are determined by three principles.
The first principle is that each son triangle inherits the material type of its father triangle.
The second principle is that,
if a new triangle overlaps a old triangle with a different material,
the matter of the overlapped area of the old triangle will be inherited by its son triangle instead of by this new triangle.
Third, for the merged triangle pair ($\bigtriangleup afe$ and $\bigtriangleup abf$), if they are overlapped by new triangles with a different material, the overlapped matters are transferred to the neighbors of the triangle pair, since they have no son triangles to inherit their matters.

\subsection{Remeshing algorithm}\label{sec.RemeshAlgorithm}

In this subsection, the four basic operations are weaved together to form an algorithm
that is sufficient to prohibit mesh distortion with finite operations.

\begin{figure}[htpb]
  \centering
  \includegraphics[width=0.5\textwidth]{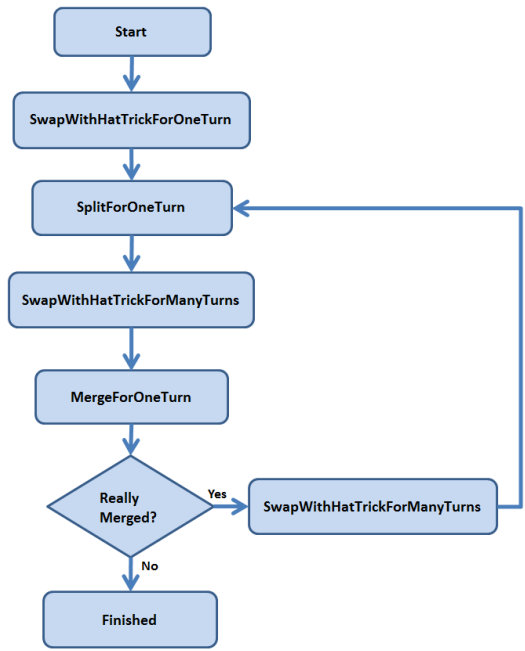}
  \caption{Main flow of the remeshing algorithm.}\label{fig.RemeshMainFlow}
\end{figure}

The main flow of the remeshing algorithm is shown in Fig. \ref{fig.RemeshMainFlow}.
``SwapWithHatTrickForOneTurn'' scans over all the triangles in the mesh for one turn,
and carries out an edge-swapping operation each time it encounters a triangle that satisfies the corresponding criterions.
If an edge-swapping operation is constrained on interface, a hat-trick operation is applied to replace it.
``SwapWithHatTrickForManyTurns'' carries out the ``SwapWithHatTrickForOneTurn'' for arbitrary times, until
no new operations are done on the mesh.
``SplitForOneTurn'' (``MergeForOneTurn'') scans over all the triangles in the mesh for one turn,
and carries out an edge-splitting (edge-merging) operation each time it encounter a triangle that satisfies the corresponding criterions.

The criterions for edge-swapping and hat-trick in ``SwapForOneTurnWithHatTrick'' are described by the following algorithm.

\scriptsize
\begin{spacing}{0.8} % for 0.8 normal line spacing
\begin{verbatim}
if(cosine of the largest angle of the triangle < -0.5)
{
  find the triangle that is neighbored to its largest edge
  if(the triangle and the neighbored triangle contain the same material)
  {
    do edge-swapping on this triangle pair
  }
  else
  {
    if(cosine of the largest angle of the triangle < -0.7)
    {
      if (triangles surrounding the vertex of largest angle contain the same material)
      {
        do hat-trick on this triangle
      }
      else
      {
        do edge-swapping on this triangle pair
      }
    }
  }
}
\end{verbatim}
\end{spacing}
\normalsize

The criterions for edge-splitting operation in ``SplitForOneTurn'' are described by the following algorithm.

\scriptsize
\begin{spacing}{0.8} % for 0.8 normal line spacing
\begin{verbatim}
foreach(edge of the triangle's edges)
{
  find the triangle that is neighbored to this edge
  if(the triangle and the neighbored triangle contain the same material)
  {
    if(edge length > 2.0 * StandardLength)
    {
      do edge-splitting on this edge
    }
  }
  else
  {
    if(edge length > 1.0 * StandardLength)
    {
      do edge-splitting on this edge
    }
  }
}
\end{verbatim}
\end{spacing}
\normalsize

The criterions for edge-merging operation in ``MergeForOneTurn'' are presented by the following algorithm.

\scriptsize
\begin{spacing}{0.8} % for 0.8 normal line spacing
\begin{verbatim}
find the shortest edge of the triangle
find the triangle that is neighbored to this edge
{
  if(edge length < 0.25 * StandardLength)
  {
    doMerging = true
  }
  else if(the triangle and the neighbored triangle contain the same material)
  {
    if(edge length < 0.35 * StandardLength)
    {
      doMerging = true
    }
    else if(either of the two end points of the edge is inside material)
    {
      if(edge length < 0.5 * StandardLength)
      {
        doMerging = true
      }
    }
  }
  if(doMerging is true)
  {
    N_Left = the total material types surrounding the left end point of the edge
    N_Right = the total material types surrounding the right end point of the edge
    if(N_Left <= N_Right)
    {
      do edge-merging on this edge by deleting the left end point vertex
    }
    else
    {
      do edge-merging on this edge by deleting the right end point vertex
    }
  }
}
\end{verbatim}
\end{spacing}
\normalsize

\begin{figure}[htpb]
  \centering
  \includegraphics[width=3.0in]{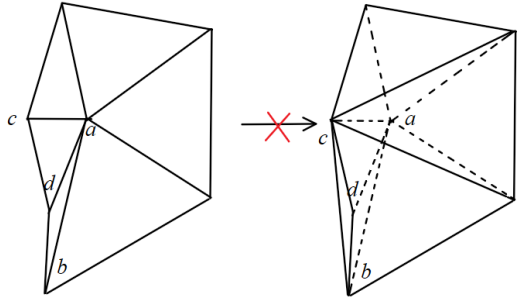}\\
  \caption{Edge-merging cancellation.}\label{fig.EdgeMergeCancel}
\end{figure}

The edge-merging operation may sometimes encounter minus overlapping coefficients.
For example, in Fig. \ref{fig.EdgeMergeCancel}, the coefficients for the new triangle $\bigtriangleup bcd$ are all minus.
In this case, the edge-merging operation is canceled.

\section{Fix of the checkerboard oscillation}\label{sec.FixOscillation}

\subsection{The description}\label{sec.CheckerboardAnaysis}

\begin{figure}[htpb]
  \centering
  \subfigure[]{\includegraphics[width=0.2\textwidth]{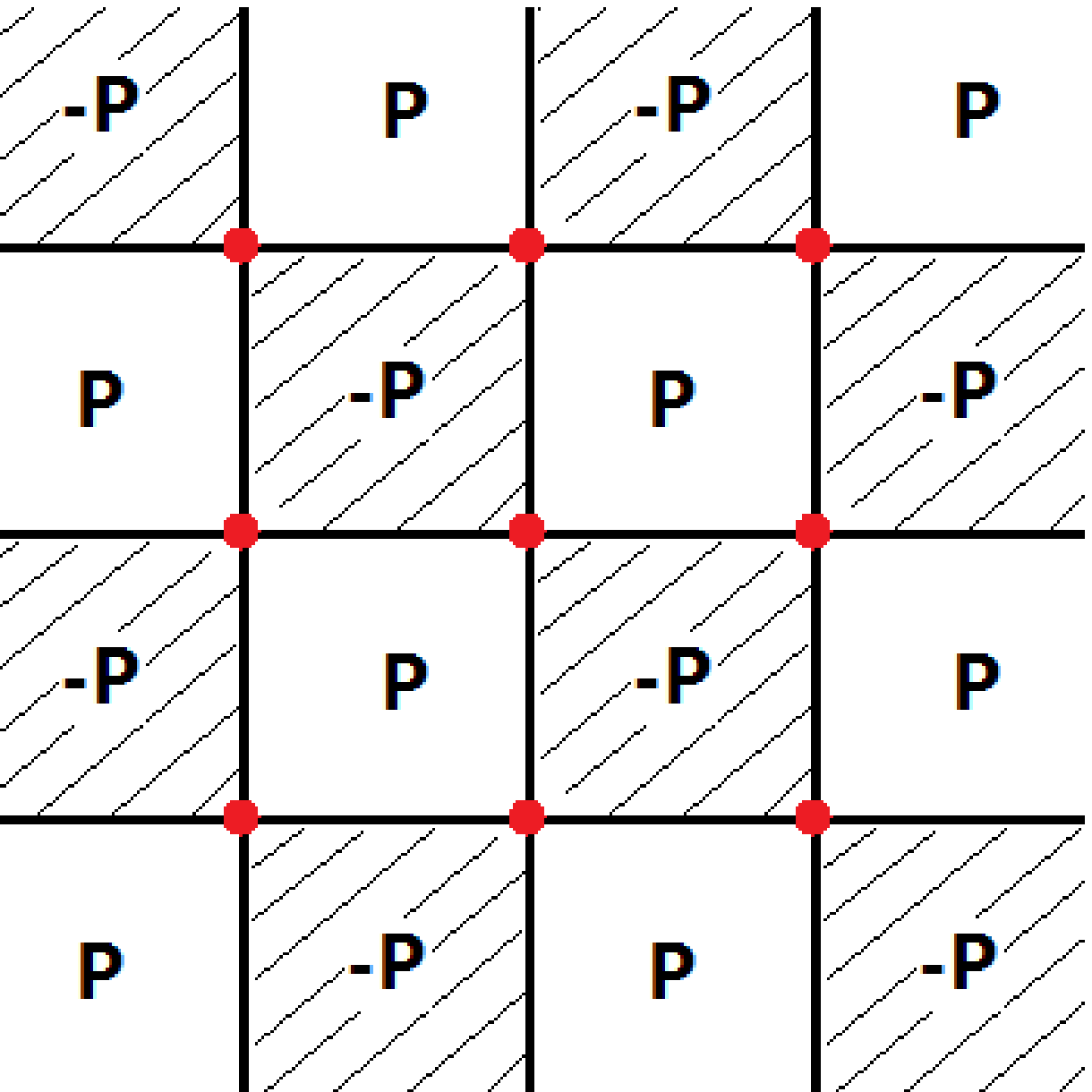}\label{fig.CheckerboardSquar}}
  \ \ \ \ \ \
  \subfigure[]{\includegraphics[width=0.2\textwidth]{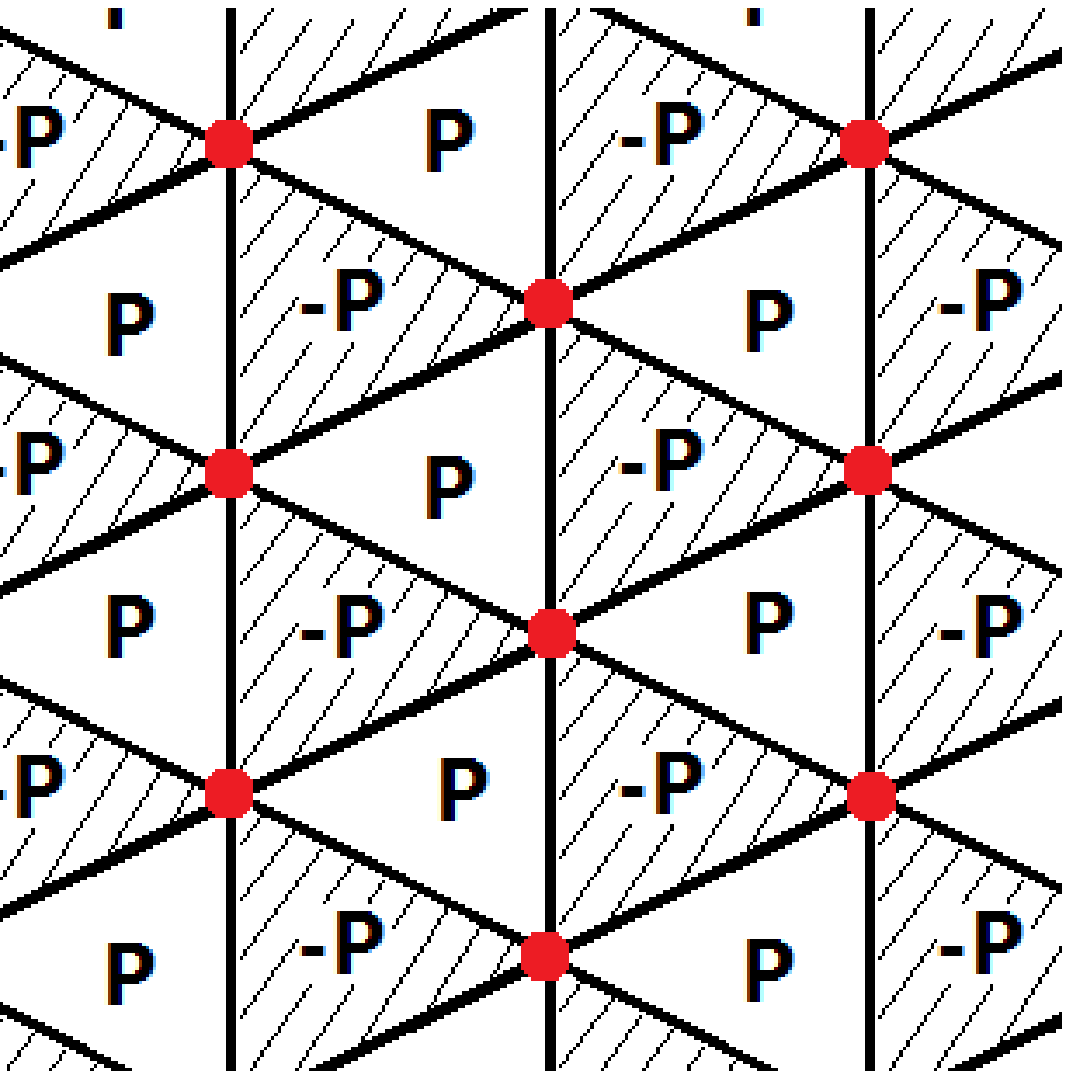}\label{fig.CheckerboardTrg}}
  \caption{Checkerboard oscillation in quadrilateral mesh and triangular mesh.}\label{fig.Checkerboard}
\end{figure}

In hydrodynamic simulations with Lagrangian frame, there exists a problem usually referred to as the ``checkerboard'' oscillation,
which corresponds to nonphysical oscillations of the pressure (or density) distributions.
Fig. \ref{fig.Checkerboard} shows two idealized examples of checkerboard oscillation on pressure.
The checkerboard pressure corresponds to an equilibrium state in Lagrangian scheme since each vertex feels a zero resultant force,
while it is not a equilibrium state in physical truth or in Eulerian simulation since matter will flow from the higher pressure cell to
the lower one.
The checkerboard oscillation problem can be more precisely stated as:
an arbitrary checkerboard pressure (density) distribution could be add to the true physical solution,
which does not alter the evolution in Lagrangian simulations.

\begin{figure}[htpb]
  \centering
  \subfigure[]{\includegraphics[width=0.25\textwidth]{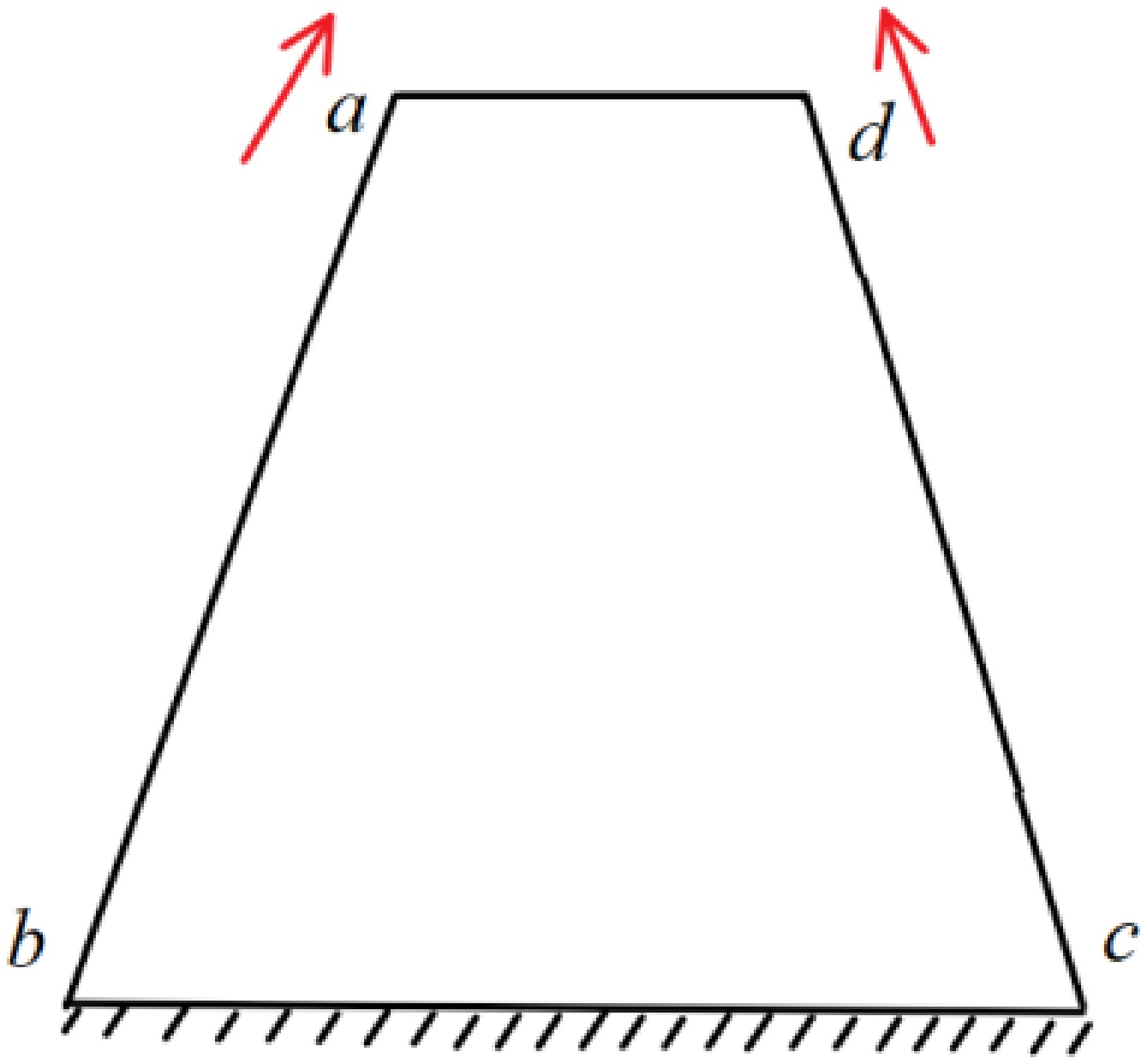}\label{fig.NonPhysicalTrgSingular1}}
  \subfigure[]{\includegraphics[width=0.25\textwidth]{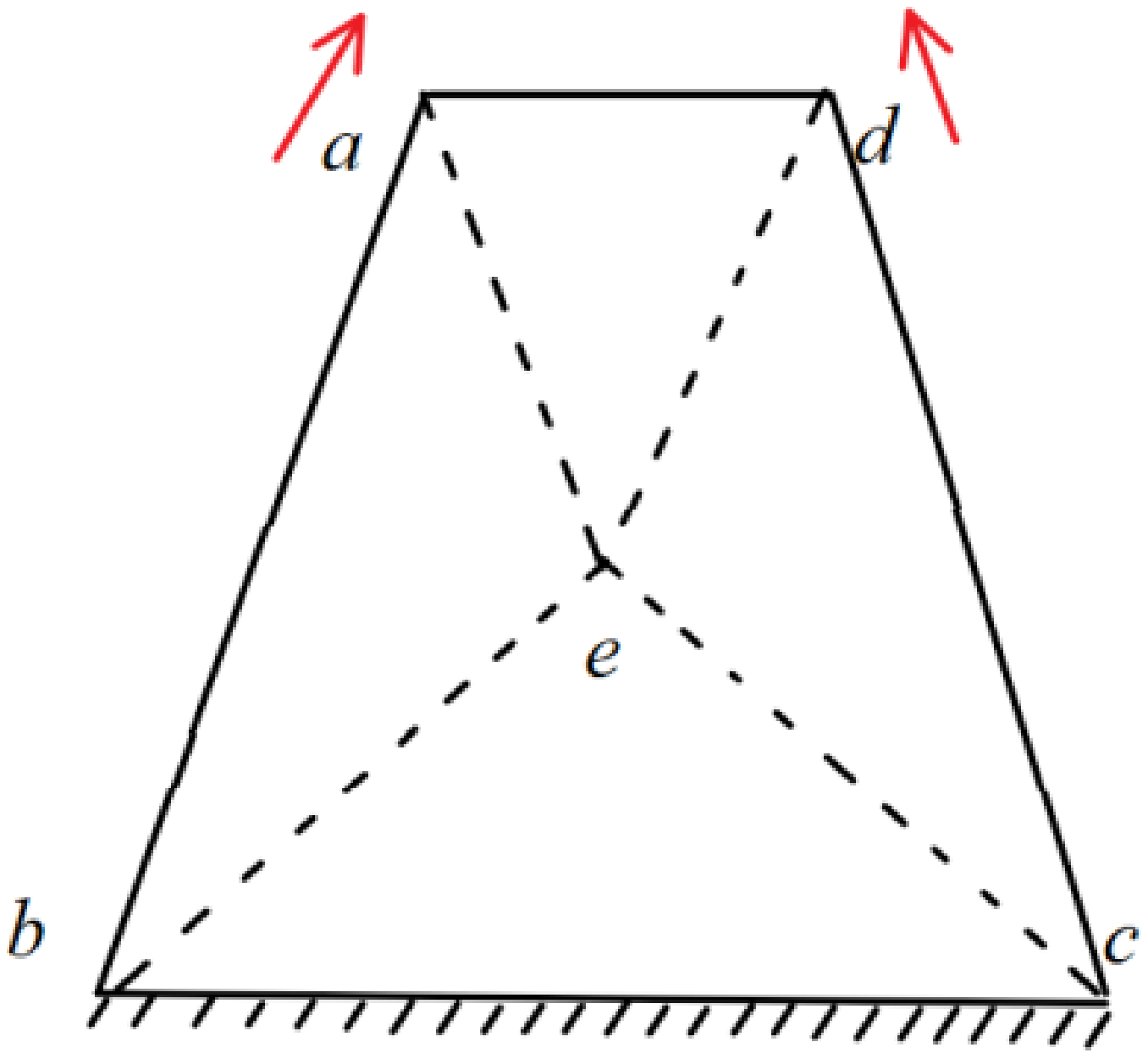}\label{fig.NonPhysicalTrgSingular2}}
  \subfigure[]{\includegraphics[width=0.25\textwidth]{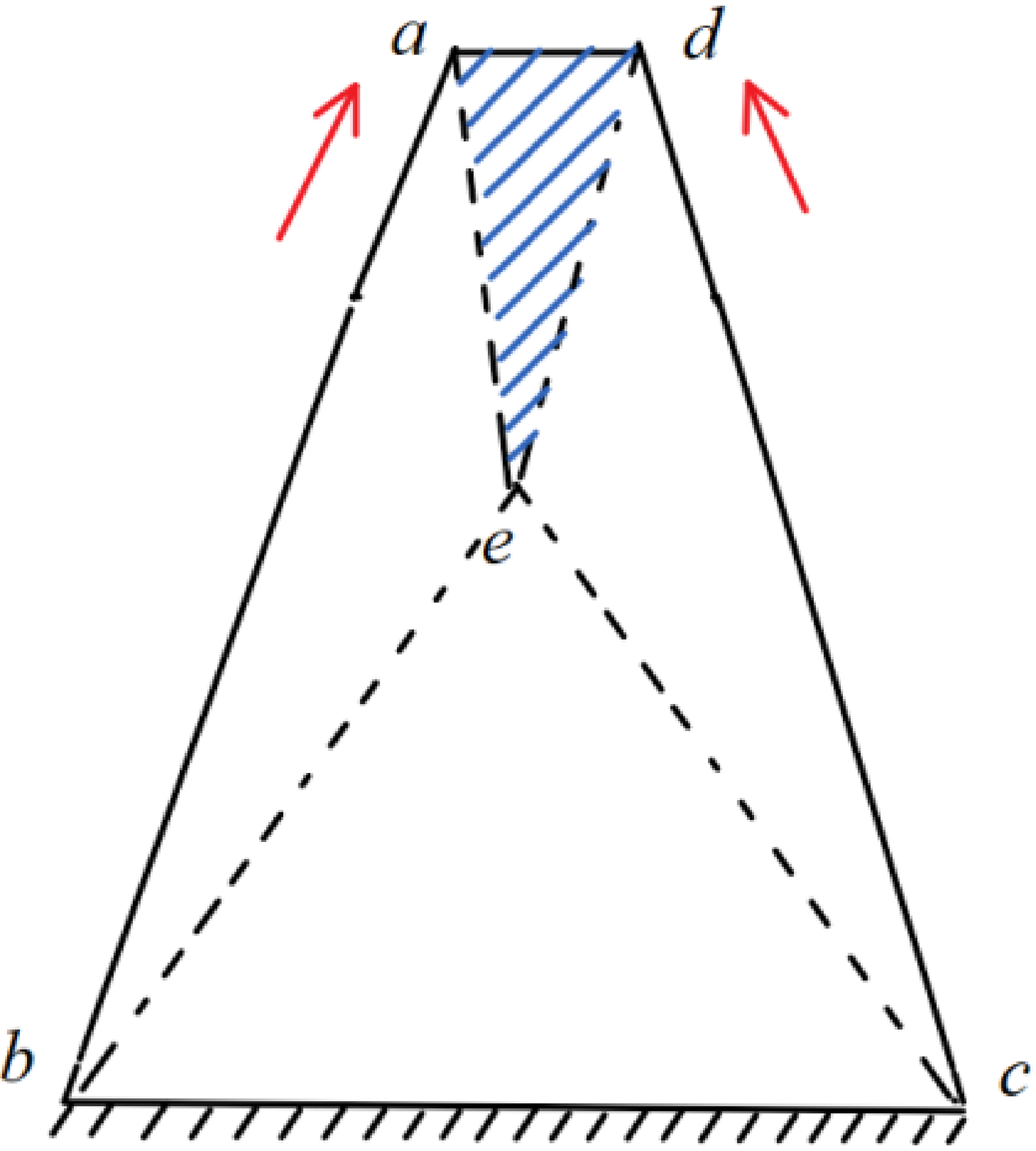}\label{fig.NonPhysicalTrgSingular3}}
  \caption{Nonphysical singular movement of triangle \cite{Caramana:1998}.}\label{fig.NonPhysicalTrgSingular}
\end{figure}

Even though a checkerboard distribution is easily to be constructed by hand,
it is not mandatory to emerge in Lagrangian simulations.
Usually, the checkerboard oscillation is easy to emerge for triangular mesh but not so frequently encountered for other polygon mesh.
This may be explained in some degree by the example given in Fig. \ref{fig.NonPhysicalTrgSingular} \cite{Caramana:1998}. Fig. \ref{fig.NonPhysicalTrgSingular1} use the quadrilateral $\square abcd$ to simulate the flow of the fluid, while Fig. \ref{fig.NonPhysicalTrgSingular2}-\ref{fig.NonPhysicalTrgSingular3} use the triangle grids. The arrows indicate the moving direction. The density and the pressure of the quadrilateral $\square abcd$ decrease as the vertices move in Fig. \ref{fig.NonPhysicalTrgSingular1}. But in Fig. \ref{fig.NonPhysicalTrgSingular3}), the area of the triangle $\bigtriangleup ade$ goes to nearly zero and gives rise to a nonphysical singularity. This example reveals that, in comparison to other polygons, triangle lacks the degree of the freedom to fully simulate the flow.

\subsection{The compensation for mesh-bending}

Inspired partly by the analysis in Sec. \ref{sec.CheckerboardAnaysis}, we trace the root of the checkerboard oscillation
back to the non-bending of the grid edges.
Taking again the Fig. \ref{fig.Checkerboard} for example,
in a Lagrangian view, it's true that the vertices do not need to move, but there should be a bending movement of the edges.
The edge-bending movement can be captured by some advanced Lagrangian methods \cite{Margolin:1999,Andrew:2016}.
In this work, we try to introduce an Euler-like flow on each edge to compensate for the bending effects so as to mitigate the checkerboard oscillation.

\begin{figure}[htpb]
  \centering
  \subfigure[]{\includegraphics[width=0.3\textwidth]{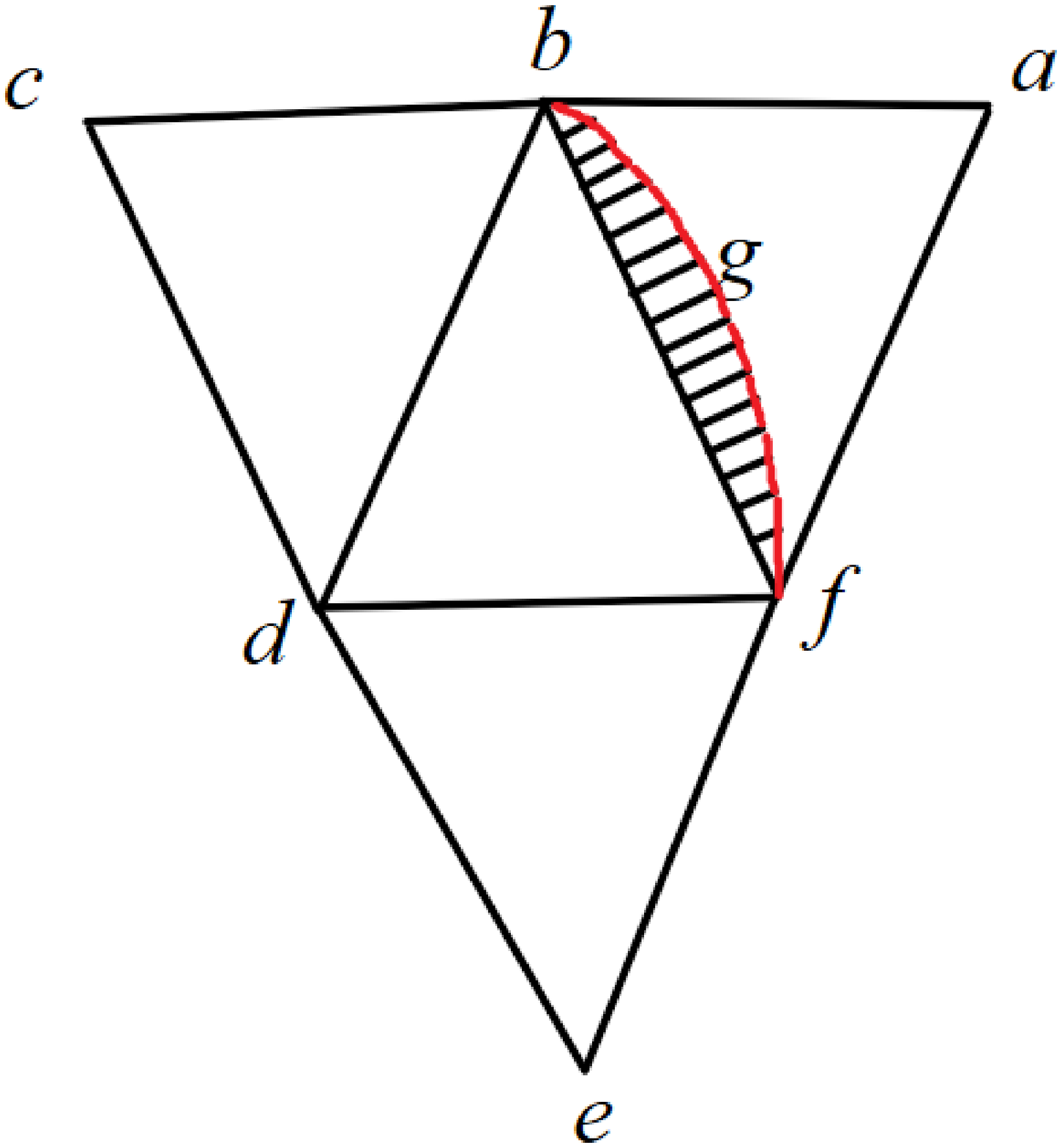}\label{fig.EdgeBent}}
  \subfigure[]{\includegraphics[width=0.3\textwidth]{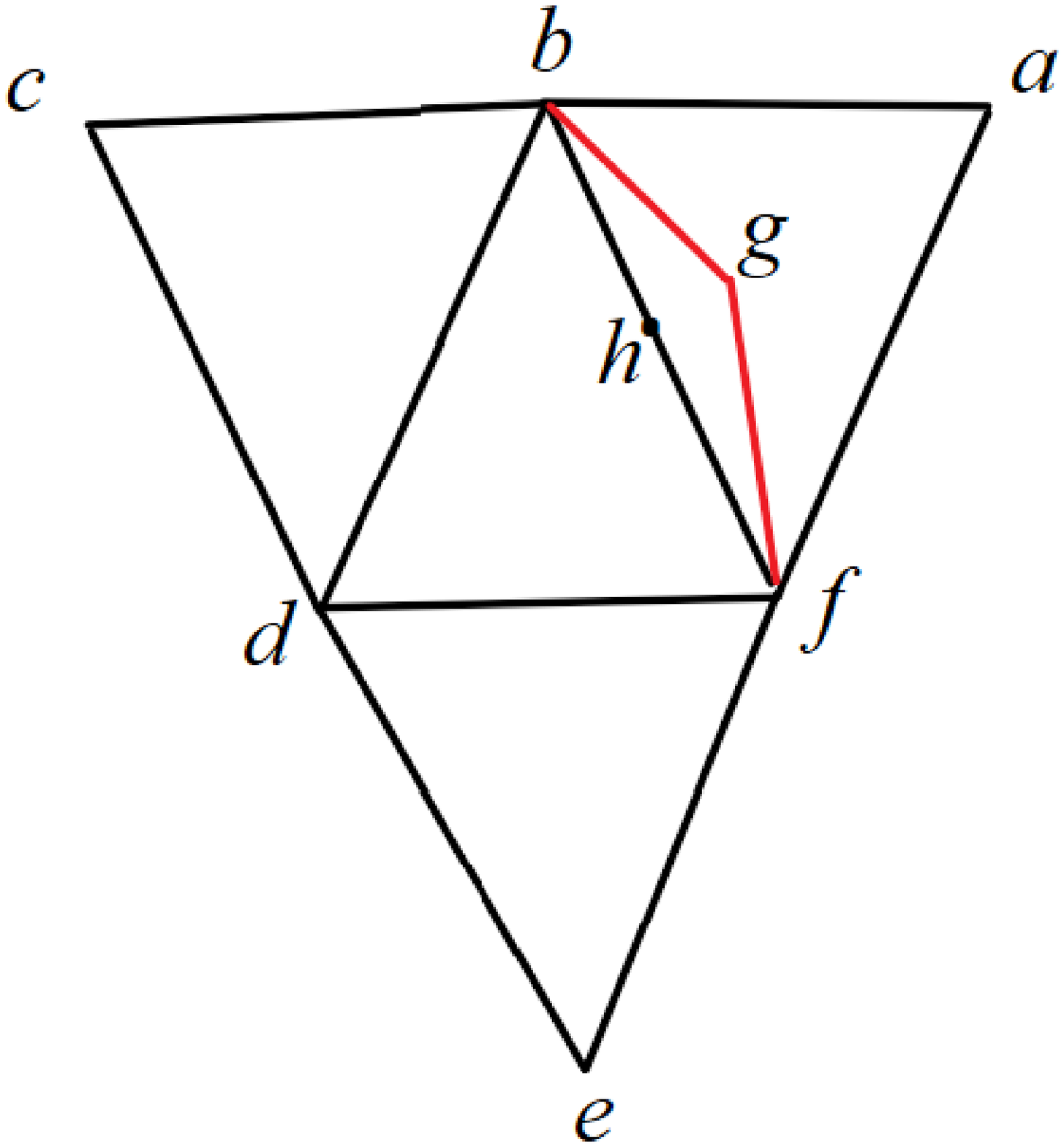}\label{fig.EdgeBentByVertexMove}}
  \caption{Edge bending treatment.}\label{fig.EdgeBentDeal}
\end{figure}

In Fig. \ref{fig.EdgeBent}, the triangle $\bigtriangleup bdf$ has higher pressure than $\bigtriangleup abf$,
which will lead to a curved edge \underline{bgf} at the next time step.
Since in usual Lagrangian simulation a straight line \underline{bf} is always kept,
an Euler-like matter flow is constructed for the edge \underline{bf} to compensate for the bending effects.
The amount of matter flow is determined by the area swept by the curve line, as shown by the shadow region in \ref{fig.EdgeBent}.
However, the real curved edge \underline{bgf} is hard to be obtained, thus a polygonal line is utilized to represent it, as shown in Fig. \ref{fig.EdgeBentByVertexMove}.
The polygonal line can be determined by imagining a vertex is inserted at the midpoint of the edge \underline{bf} and moves under the pressure.
The explicit algorithm for the matter flow compensation are as follows.
\begin{description}
  \item[1)] Calculate the average acceleration $\overline{a}$ of the midpoint $h$ on the edge \underline{bf} by
       \begin{equation}
         \overline{a}= \vec{s}_n \cdot (\vec{a}_b+\vec{a}_f)/2,
       \end{equation}
       where $\vec{a}_b$ and $\vec{a}_f$ are the accelerations of vertices $b$ and $f$ calculated by the usual Lagrangian scheme,
       and $\vec{s}_n$ is the direction vector of the edge \underline{bf}.
  \item[2)] Use the pressure difference between $\bigtriangleup bdf$ and $\bigtriangleup abf$  to obtain the acceleration $a_c$ of the imagined inserted vertex as
       \begin{equation}
        a_c = (P_{bdf} - P_{bfa}) / M_c,
       \end{equation}
      where $P_{bdf}$ and $P_{bfa}$ are the pressures of the two triangles, and $M_c$ is the mass of the vertex.
  \item[3)] The difference of $a_c$ and $\overline{a}$ decides the flow velocity $v_c$ and the area $S_{bfg}$ of the triangle $\bigtriangleup bfg$ by
       \begin{equation}
        \begin{split}
        \frac{\Delta v_c}{\Delta t} &= a_c - \overline{a}\\
        S_{bfg} &= 0.5 \times (v_c + 0.5 \times (a_c - \overline{a}) \times \Delta t)\times \Delta t \times l_{bf},
        \end{split}
       \end{equation}
      where $l_{bf}$ is the length of the edge \underline{bf}.
      Then, the compensation mass is calculated from the compensation area by
       \begin{equation}
        \Delta M_{bdf} = S_{bdf}\times \rho_{bfg},
       \end{equation}
      where $\rho_{bfg}$ equals to $\rho_{bdf}$ or $\rho_{bfa}$ depending on whether the compensation mass flows out or into $\bigtriangleup bdf$.
  \item[4)] Repeat (1)-(3) on the other two edges;
\end{description}

\section{Numerical tests}\label{sec.NumericalTests}

Four test problems including the Sod shock tube, Noh, Triple-point, and double fluid convection
are simulated in this section to assess the performance of our methods.
There are no large deformation in the Sod shock tube and Noh problems,
and so the remeshing is not introduced in those simulations,
the major aim of those tests is to assess the performance of the compensation matter flow method.
For the triple-point and double fluid convection problems,
remeshing are always carried out to deal with the fluids large deformation.

\subsection{Sod shock tube}

The domain of the tube is set as $x\in [-1,1]$, with $x=0$ separating the higher and lower pressure regions.
The initial values are given as following:
\begin{equation}
  \begin{split}
  (\rho, u, p, \gamma)_l &= (1,0,1,1.4)\ \ \ \ \text{if}\ \ -1\le x \le 0;\\
  (\rho, u, p, \gamma)_r &= (0.125,0,0.1,1.4)\ \ \ \ \text{if}\ \ 0\le x \le 1;
  \end{split}
\end{equation}
We construct a triangular mesh with $60\times 120\times 2$ elements (the factor 2 comes from that a quadrilateral splits into two triangles), as shown in Fig. \ref{fig.SodInitialMesh}.
One thing deserving to be noted is that, because the mesh is intentionally set as irregular in order to best reflect the performance of the method,
the shockwave front at the initial state is not a perfect straight line.

\begin{figure}[htpb]
  \centering
  \subfigure[]{\includegraphics[width=0.48\textwidth]{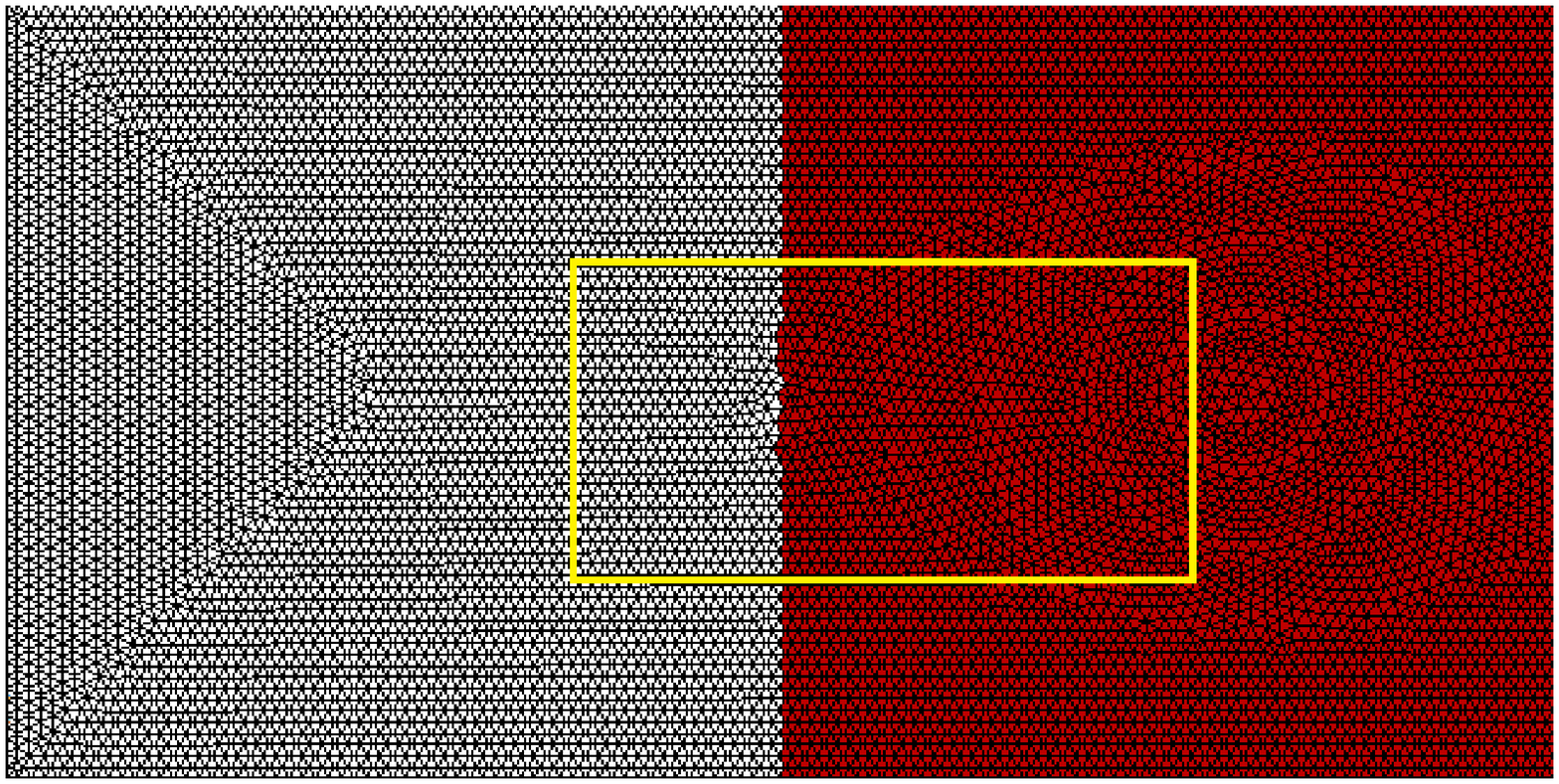}\label{fig.SodInitialMesh}}
  \subfigure[]{\includegraphics[width=0.48\textwidth]{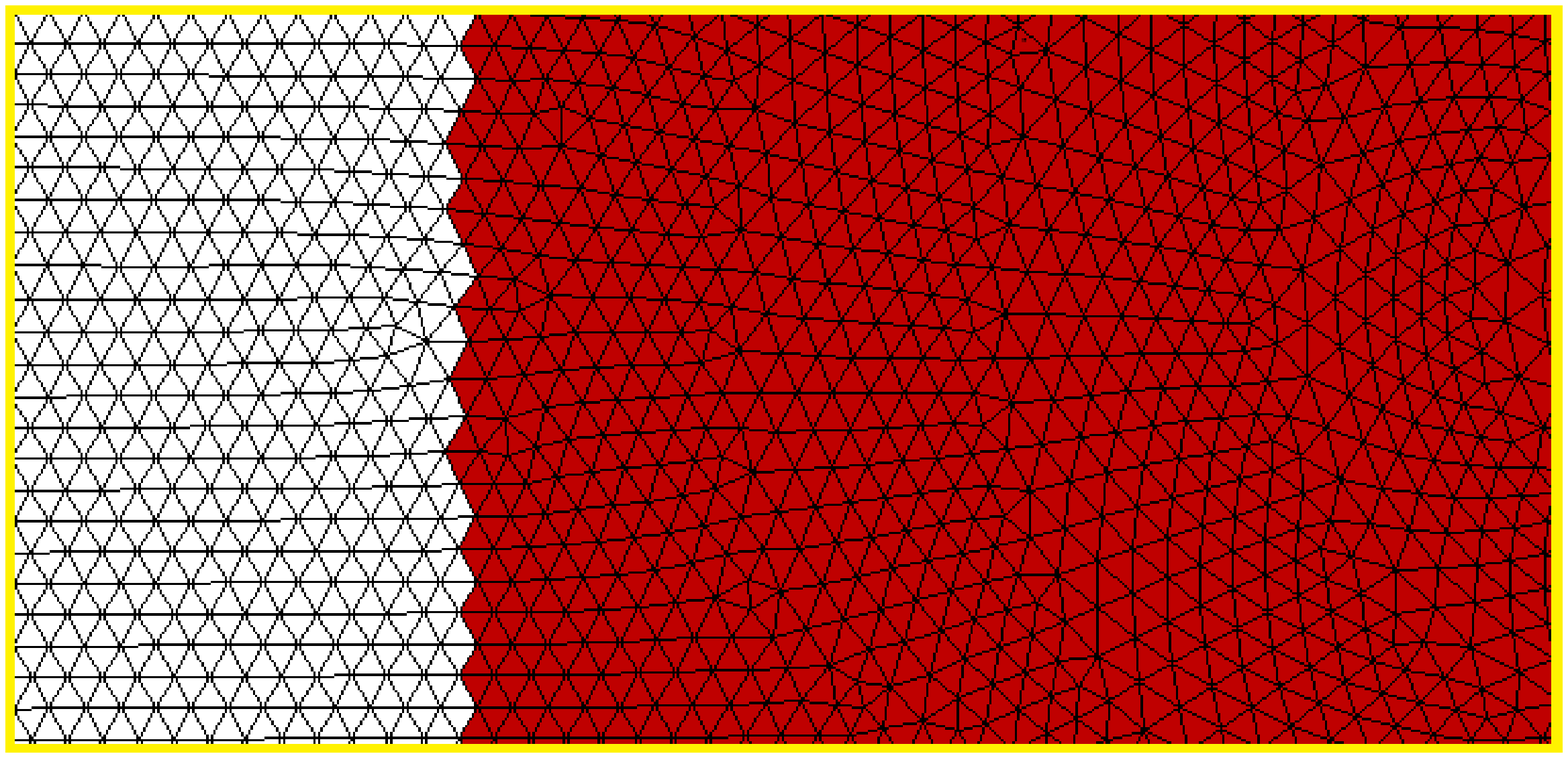}\label{fig.SodInitialMesh}}\\
  \caption{Initial mesh for the Sod simulation. The right figure is a zoom in view of the left figure.}\label{fig.SodInitialMesh}
\end{figure}

\begin{figure}[htpb]
  \centering
  \subfigure[]{\includegraphics[width=0.45\textwidth]{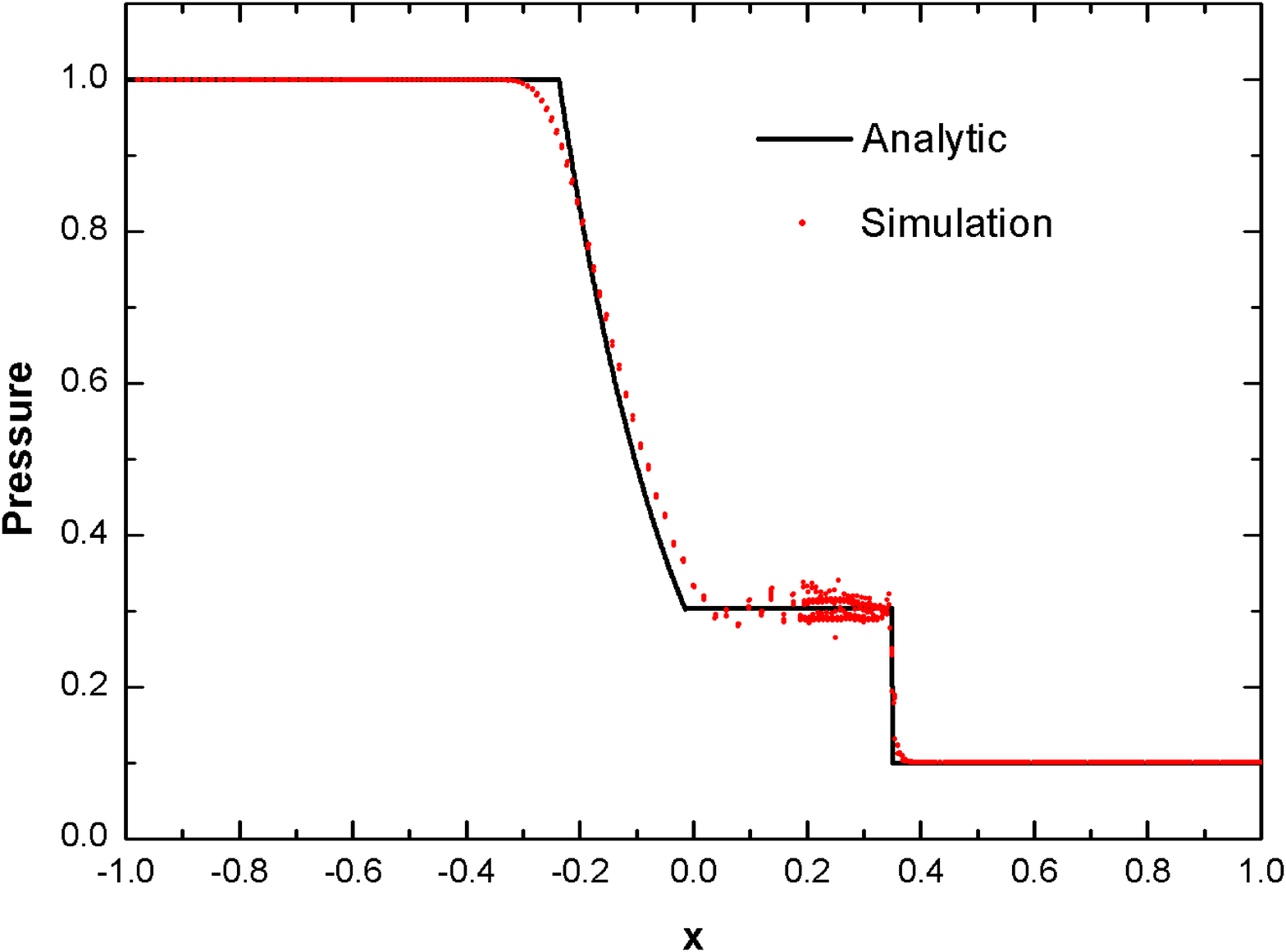}\label{fig.SodPressureNoFlow}}
  \subfigure[]{\includegraphics[width=0.45\textwidth]{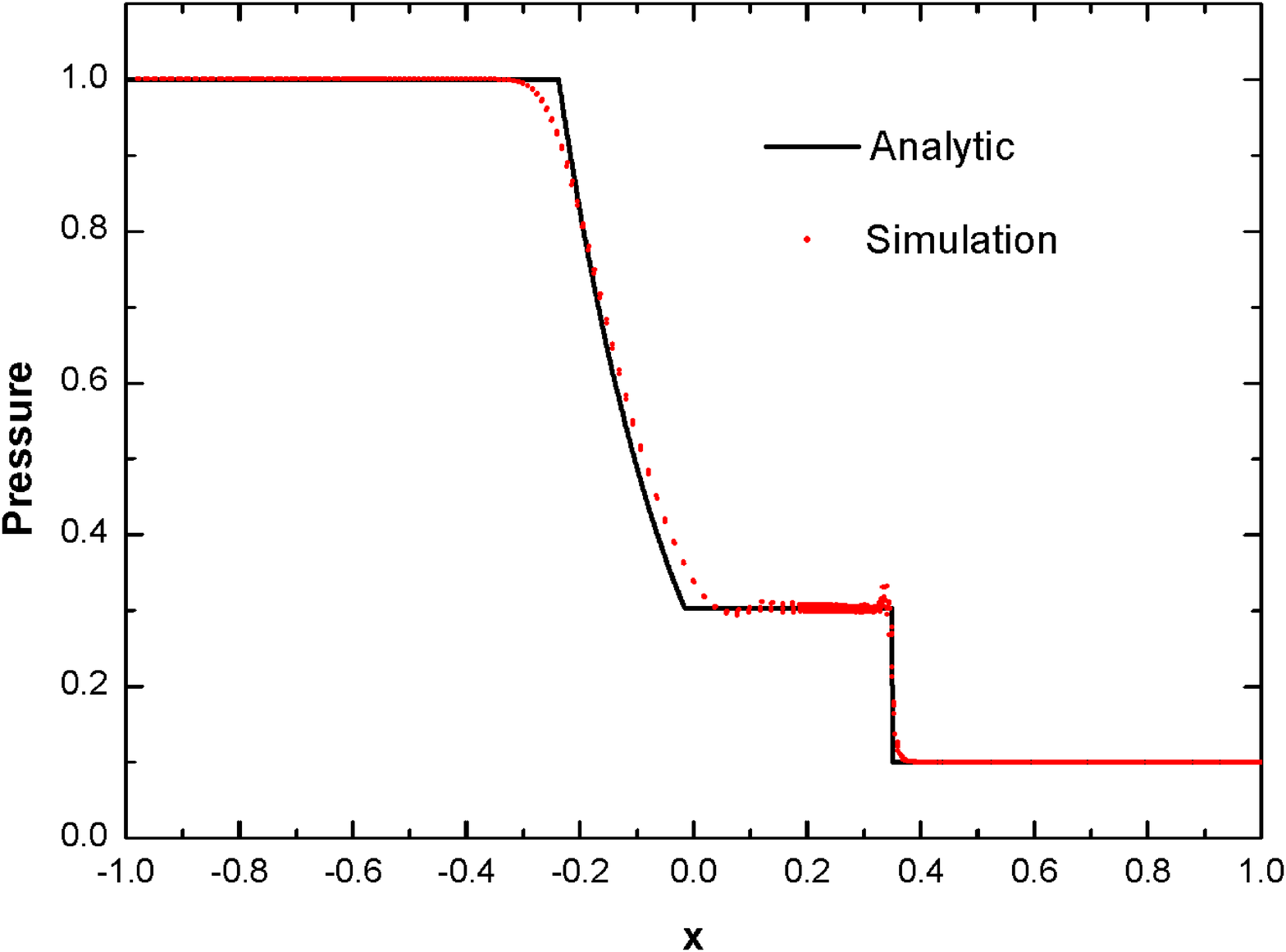}\label{fig.SodPressureWithFlow}}\\
  \caption{Pressure distributions for sod problem.
  (a) without compensation matter flow;
  (b) with compensation matter flow;
  }\label{fig.SodPressure}
\end{figure}

\begin{figure}[htpb]
  \centering
  \subfigure[]{\includegraphics[width=0.45\textwidth]{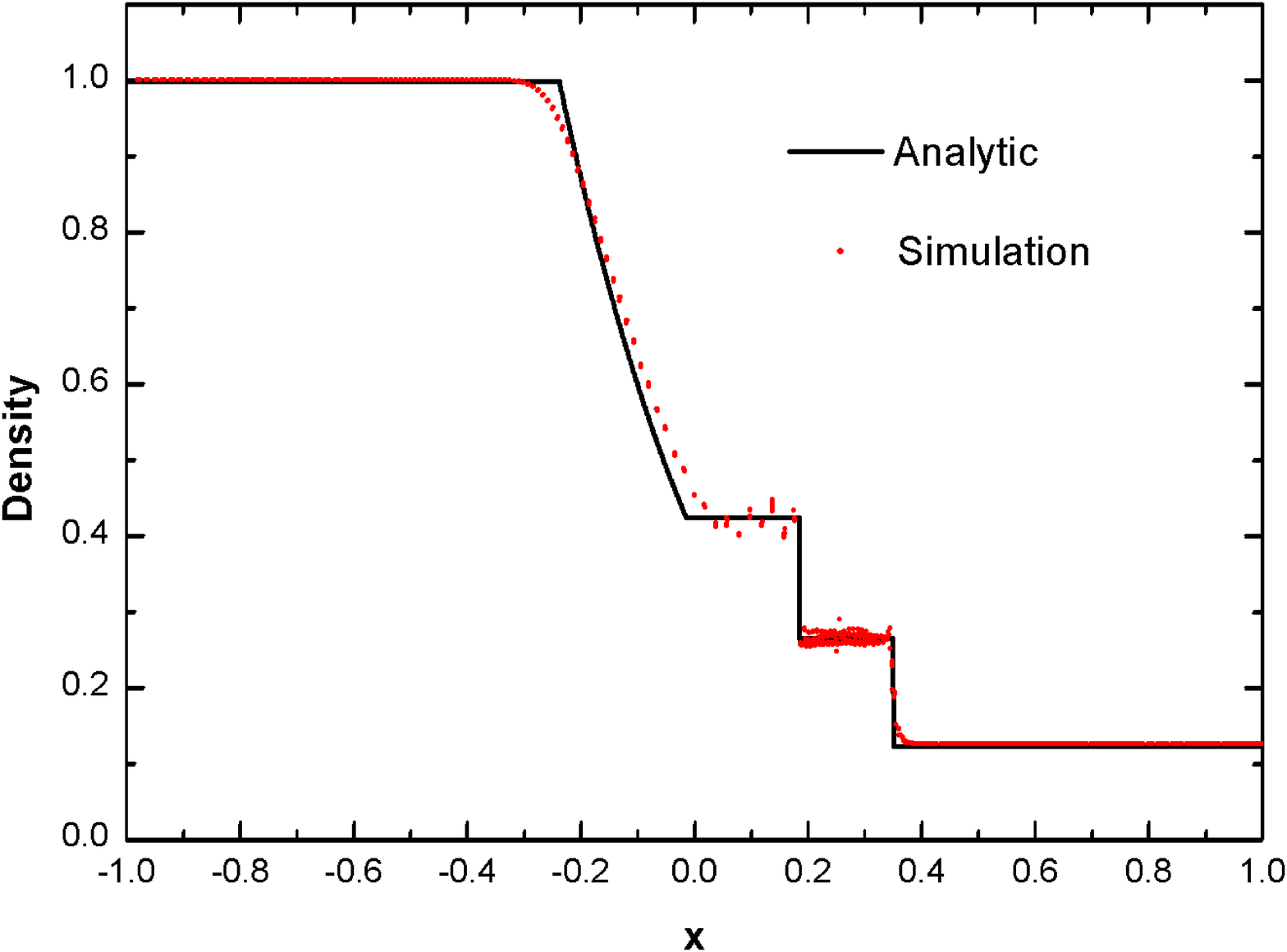}\label{fig.SodDensityNoFlow}}
  \subfigure[]{\includegraphics[width=0.45\textwidth]{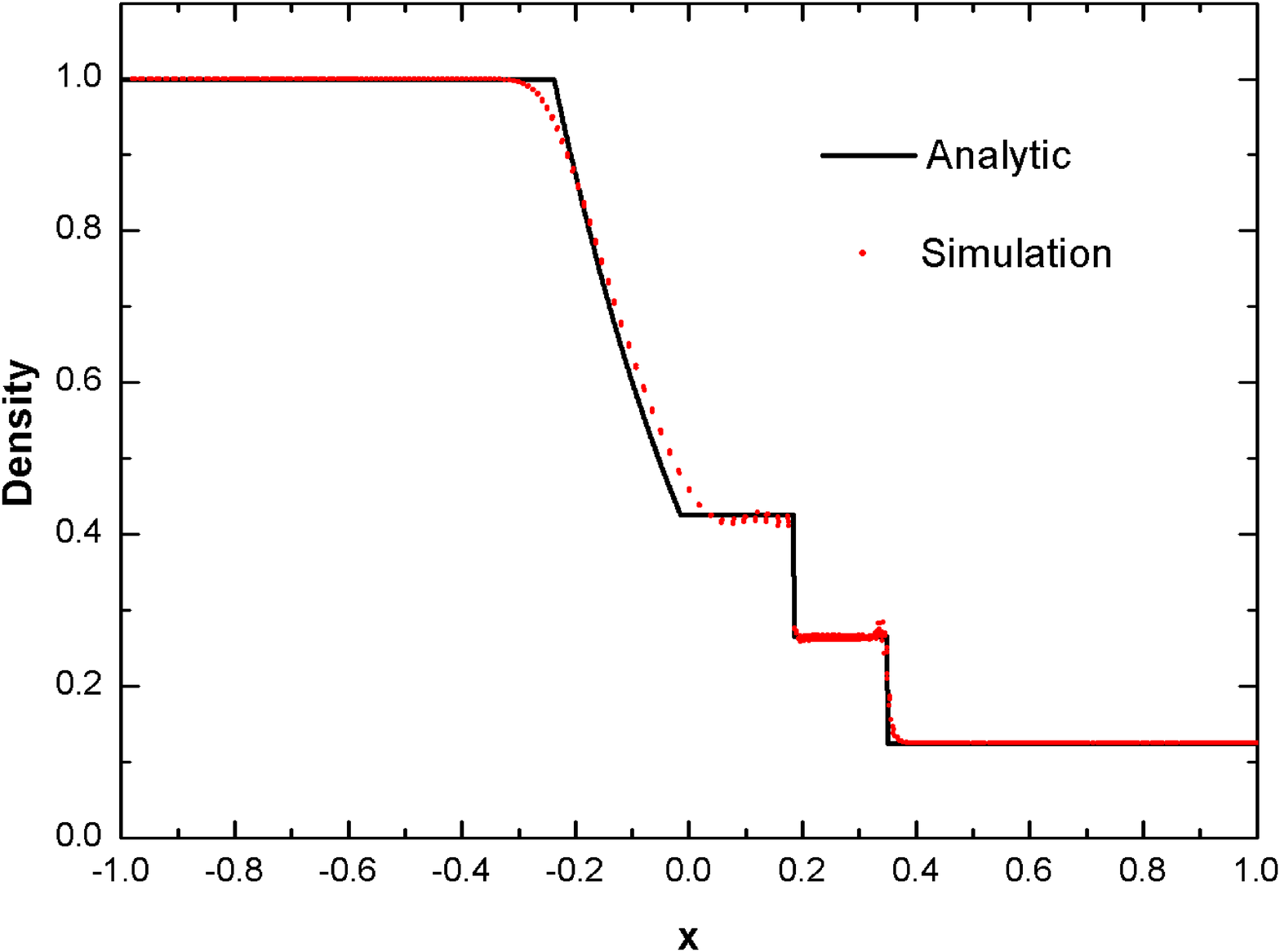}\label{fig.SodDensityWithFlow}}\\
  \caption{Density distributions for sod problem.
  (a) without compensation matter flow;
  (b) with compensation matter flow;
  }\label{fig.SodDensity}
\end{figure}

Simulation results of the density and pressure distributions with and without the compensation matter flow
are comparably shown in Figs. \ref{fig.SodPressure} and \ref{fig.SodDensity}, along with the theoretical exact solution.
It's seen that the numerical solution agrees well with the exact one,
and the compensation matter flow reduces remarkably the oscillations on the density and pressure distributions.

\subsection{Noh test}

We consider the canonical Noh \cite{Noh:1987} explosion test problem on Cartesian grid.
The problem domain consists of an ideal gas with $\gamma=5/3$ and an initial density
$\rho_0=1$ and velocity $\vec{v} = \left[-\frac{x}{\sqrt{x^2+y^2}},-\frac{y}{\sqrt{x^2+y^2}}\right]$.
This so-called Noh problem generates a expanding cylindrical shockwave from the domain center.
The exact solution is given as a function of radius $r$ and the time $t$.
At the time $t=0.6$, the exact solution is:
\begin{equation}
(\rho, e, u_r) =
\begin{cases}
(16,\frac{1}{2},0) & \text{if  } r<0.2 \\
(1+\frac{3}{5}\frac{1}{r},0,1)   & \text{if  } r>0.2
\end{cases}
\end{equation}
and the shockwave front locates at radius $r=0.2$.

\begin{figure}[htpb]
  \centering
  \subfigure[]{\includegraphics[width=0.35\textwidth]{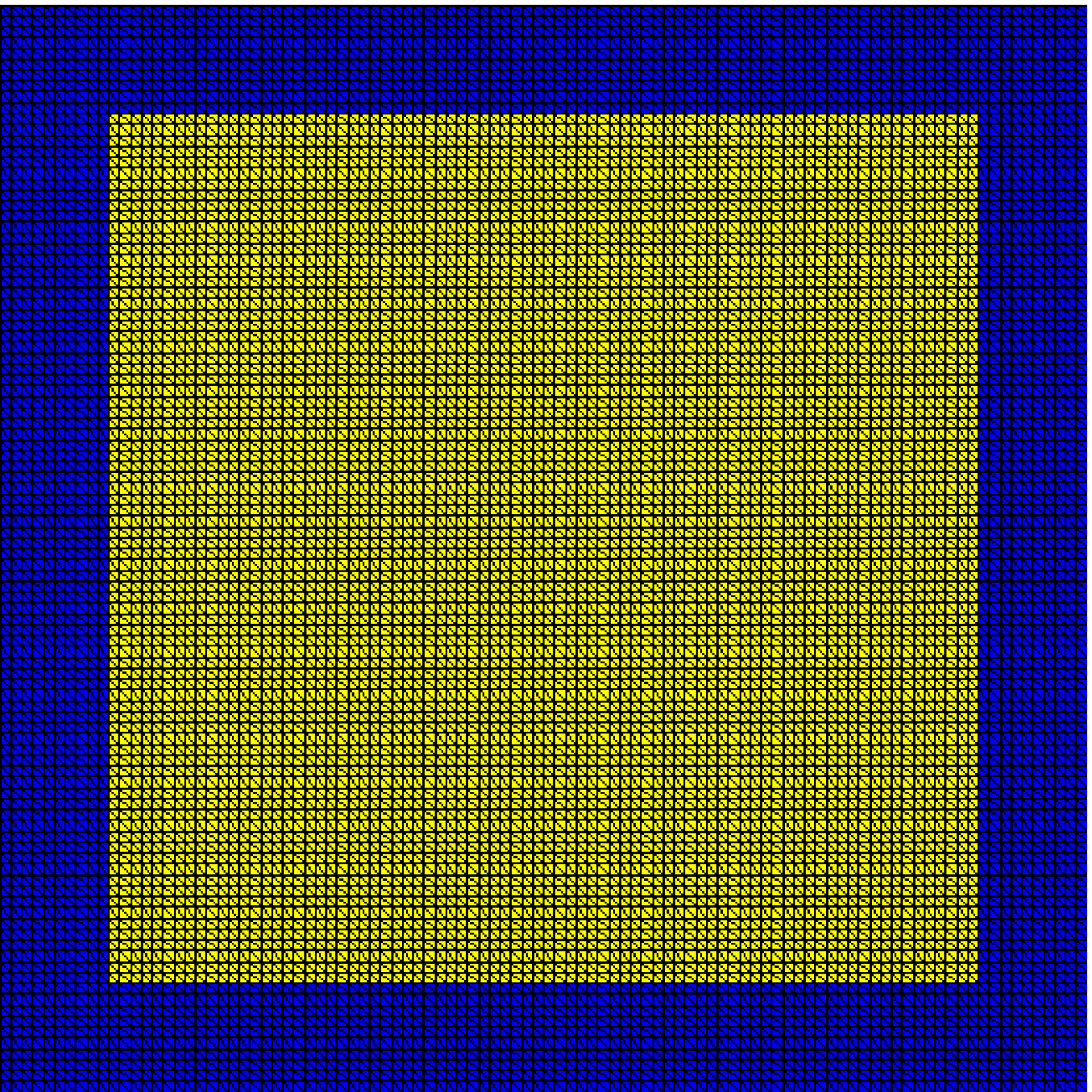}\label{fig.NohDensityPic}}
  \ \ \ \ \ \
  \subfigure[]{\includegraphics[width=0.35\textwidth]{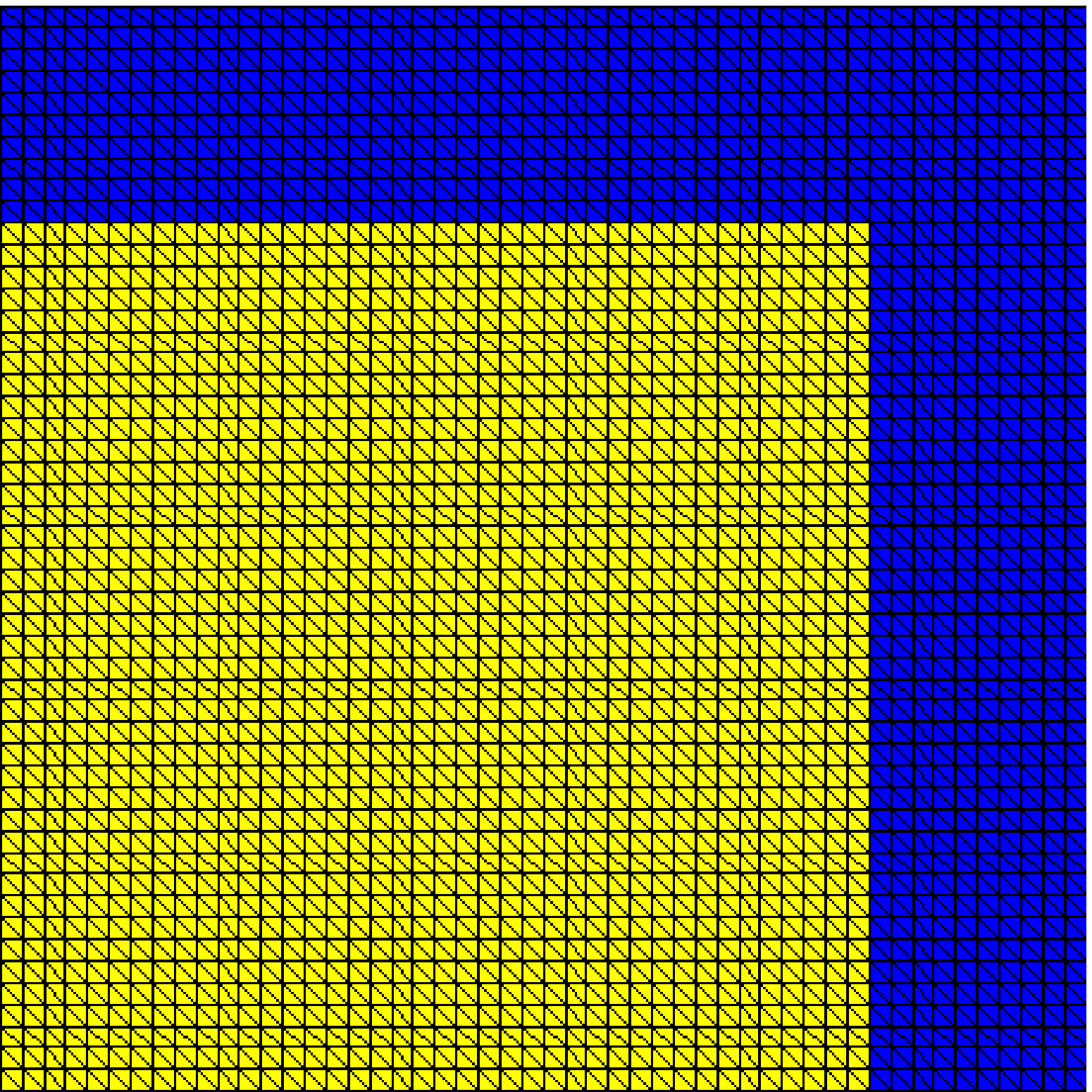}\label{fig.NohMeshPic}}\\
  \caption{Initial mesh for Noh problem simulation. The right figure is a zoom in view of the 1/4 upright part of the mesh.}\label{fig.NohMesh}
\end{figure}

We construct a triangular mesh with $100\times 100\times 2$ elements (Fig. \ref{fig.NohMesh}),
where the $80\times 80\times 2$ mesh at the center part corresponds to the matter (i.e., the ideal gas),
and the mesh on the sides is the vacuum background (represented by ideal gas with very small density).

\begin{figure}[htpb]
  \centering
  \subfigure[]{\includegraphics[width=0.3\textwidth]{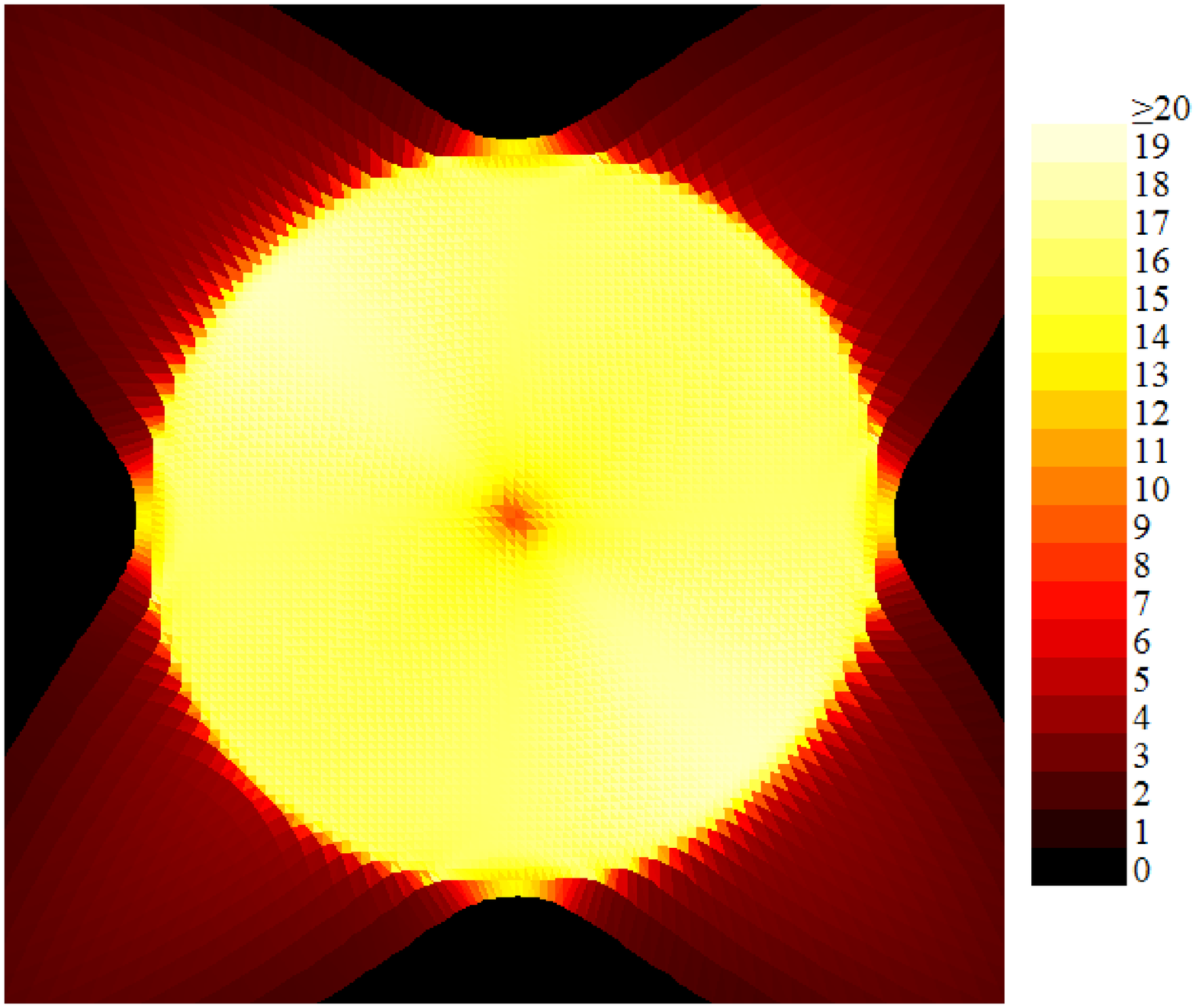}\label{fig.NohDensityPic}}
  \ \ \ \ \ \
  \subfigure[]{\includegraphics[width=0.3\textwidth]{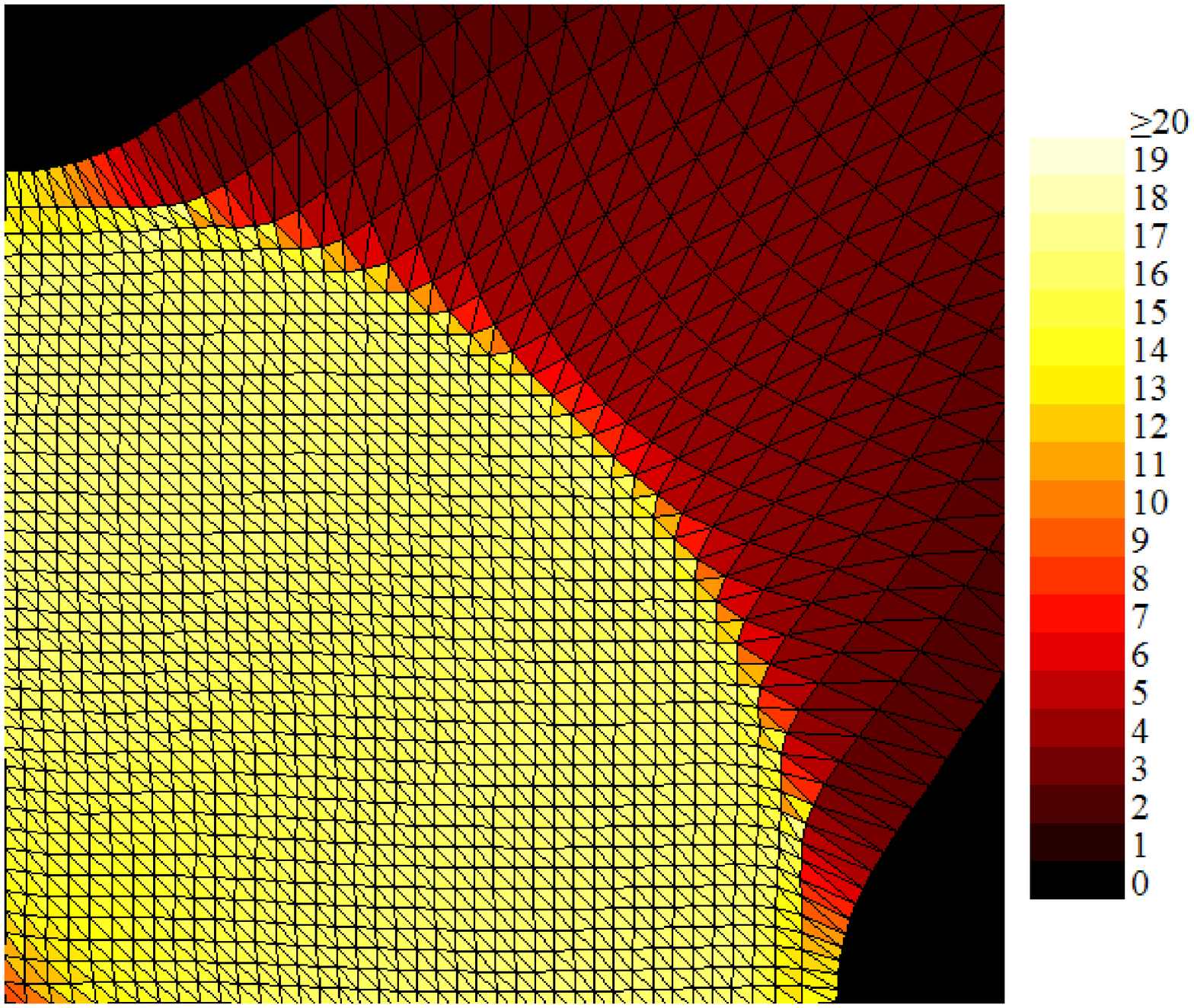}\label{fig.NohMeshPic}}\\
  \caption{(a) the density map of the domain at $t=0.6$; (b) the grids at t=0.6 (upright part).}\label{fig.NohSimuPic}
\end{figure}

\begin{figure}[htpb]
  \centering
  \subfigure[]{\includegraphics[width=0.48\textwidth]{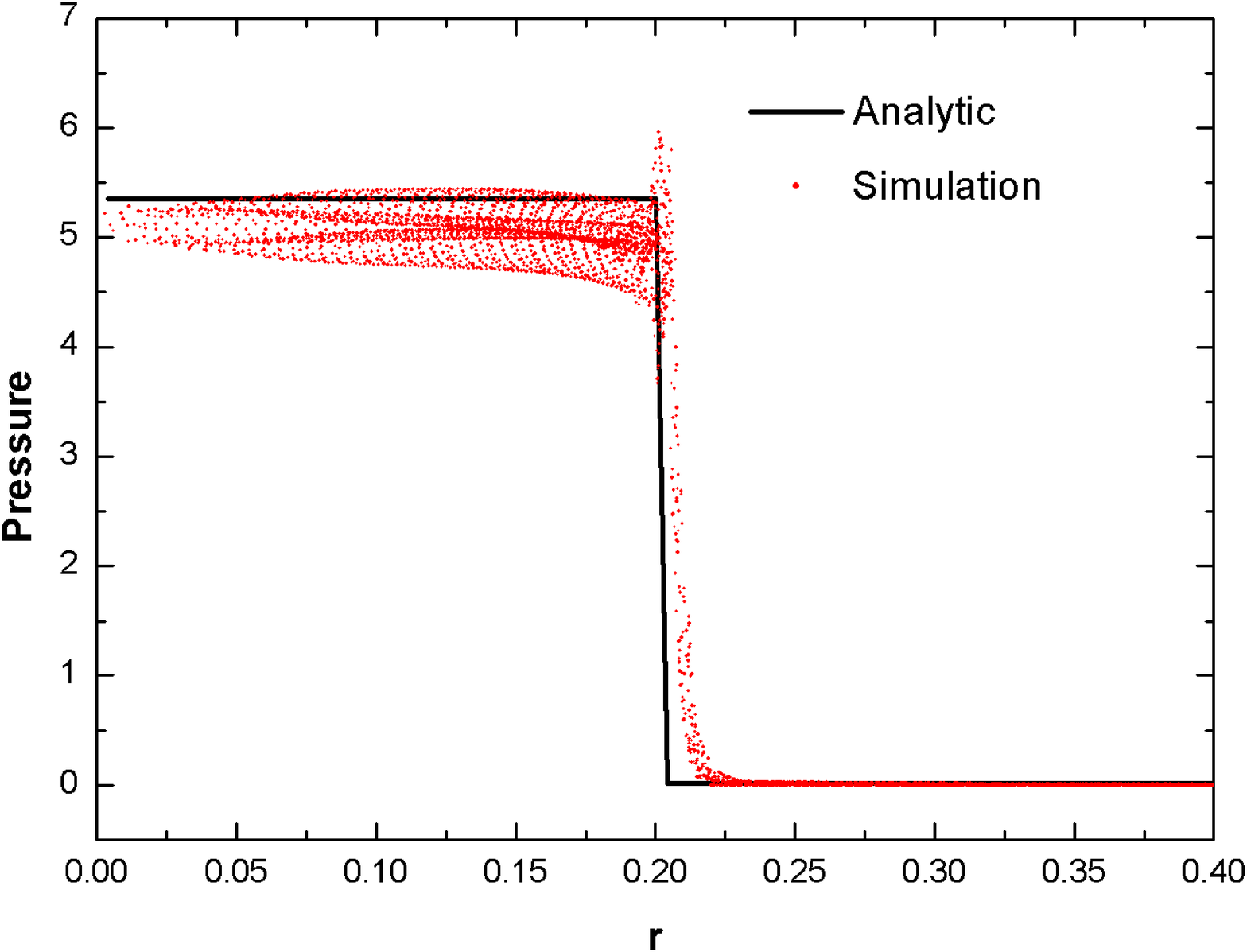}\label{fig.NohPressNoFlow}}
  \subfigure[]{\includegraphics[width=0.48\textwidth]{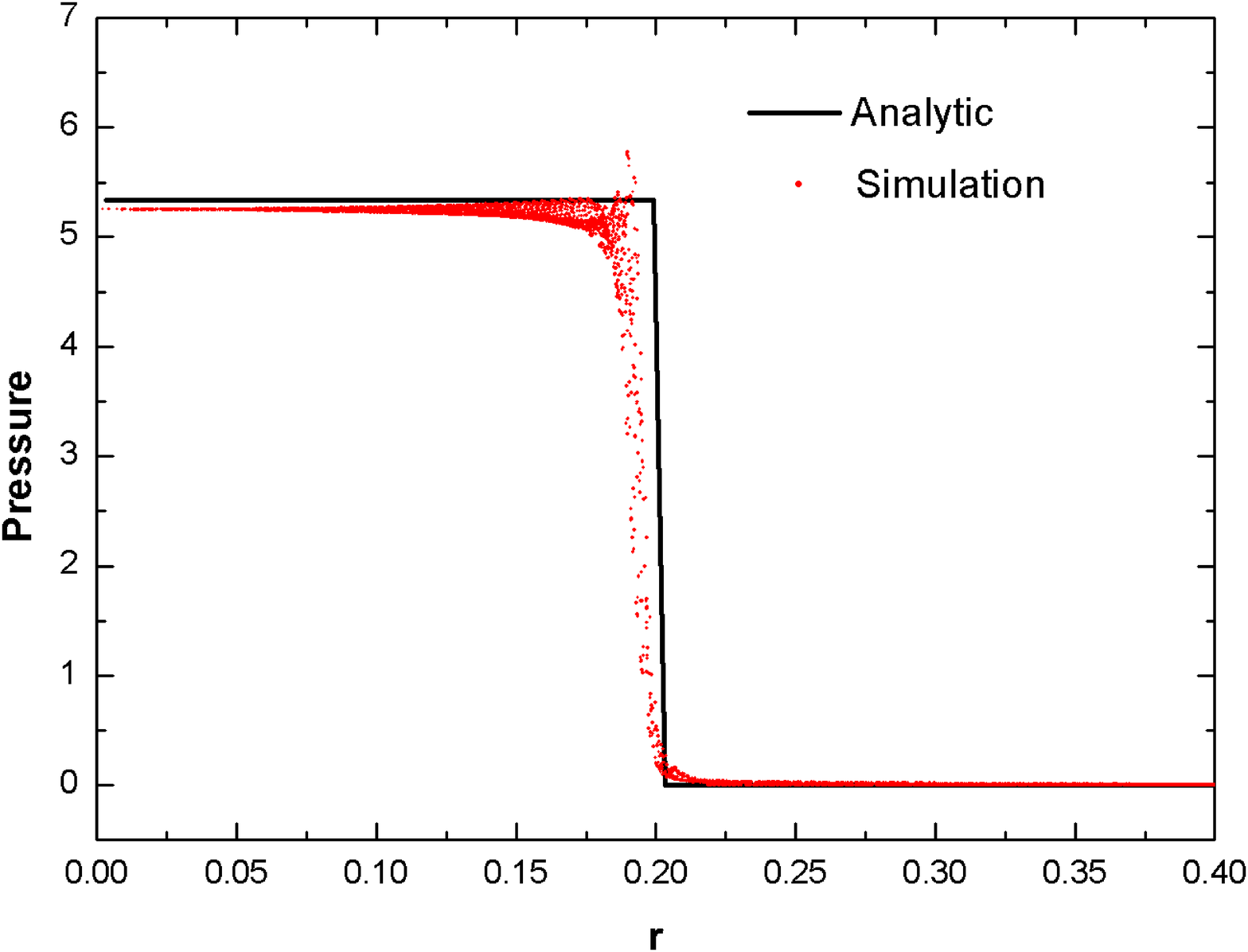}\label{fig.NohPressWithFlow}}\\
  \caption{Pressure distributions for the Noh problem at $t=0.6$. (a) without the compensation mass; (b) with the compensation mass.}\label{fig.NohPress}
\end{figure}

\begin{figure}[htpb]
  \centering
  \subfigure[]{\includegraphics[width=0.48\textwidth]{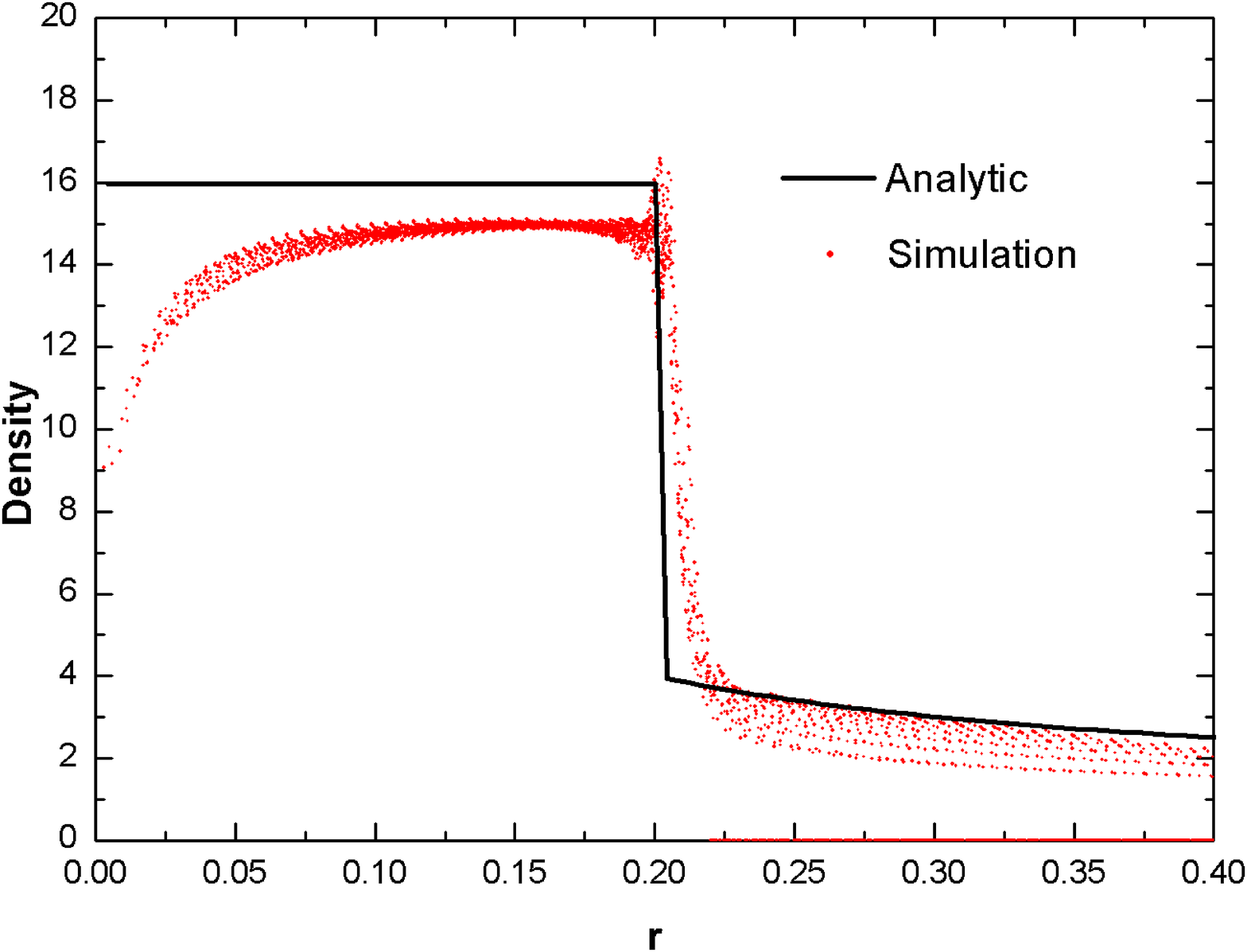}\label{fig.NohDensityNoFlow}}
  \subfigure[]{\includegraphics[width=0.48\textwidth]{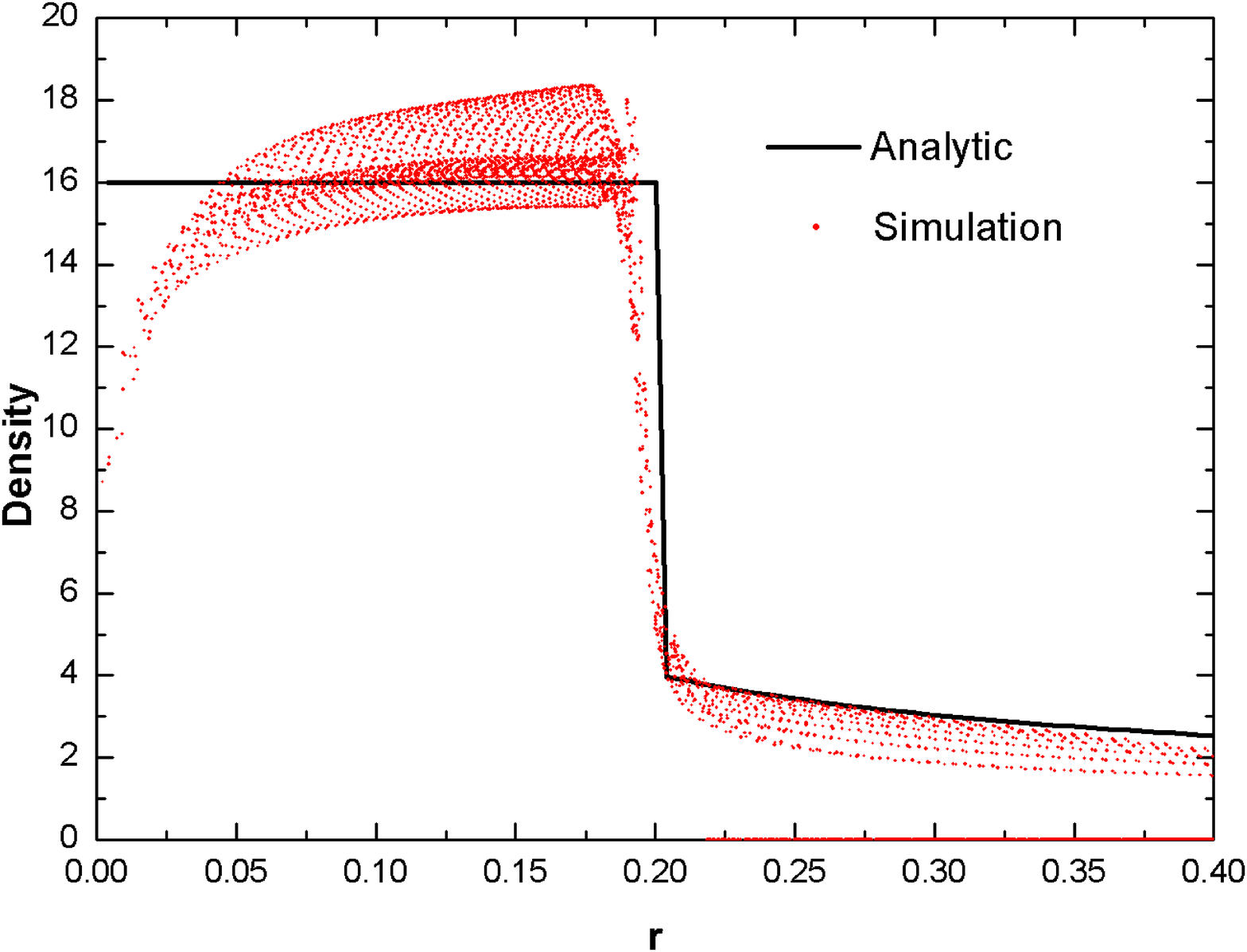}\label{fig.NohDensityWithFlow}}\\
  \caption{Density distributions for the Noh problem at $t=0.6$. (a) without the compensation mass; (b) with the compensation mass.}\label{fig.NohDensity}
\end{figure}

Fig. \ref{fig.NohSimuPic} shows the density map and the grids at $t=0.6$.
The pressure and density distribution curves at $t=0.6$ are presented in Figs. \ref{fig.NohPress} and \ref{fig.NohDensity}.
The results with and without the compensation matter flow are comparably shown in the figures, along with the analytic solutions.
For the pressure distributions, the simulation with the compensation matter flow has less oscillations than that without the compensation,
but for the density distributions the one without the compensation is smoother.
The amplitudes of the shockwave and locations of the wavefront also possess some discrepancy for simulations with and without the compensation matter flow.
Those results shows that the role the compensation matter flow plays in this simulation is complicated.
In fact, in this simulation the viscosity ($c=0.003$ is adopted here) also plays an important role that influences both the oscillations and the solutions.

\subsection{Triple-point problem}

\begin{figure}[htpb]
  \centering
  \includegraphics[width=0.6\textwidth]{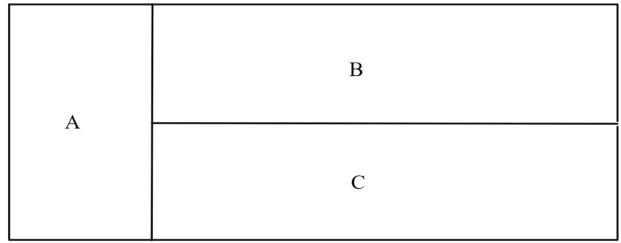}
  \caption{The triple-point problem.}\label{fig.TriplePointProblem}
\end{figure}

The triple-point problem is described in Fig. \ref{fig.TriplePointProblem}.
The computational domain $\Omega=[0,7]\times[0,3]$ splits into three sub-domains with different densities and pressures.
The initial values are given as:
\begin{equation}
  \begin{split}
  \Omega_A &= [0,1]\times[0,3]:\ \ (\rho,u,p,\gamma)=(1,0,1,1.4)\\
  \Omega_B &= [1,7]\times[1.5,3]:\ \ (\rho,u,p,\gamma)=(0.1,0,0.125,1.4)\\
  \Omega_C &= [1,7]\times[0,1.5]:\ \ (\rho,u,p,\gamma)=(1,0,0.125,1.4)
  \end{split}
\end{equation}

The triple-point problem can be simulated by two kinds of settings: one is adopting the same material
in the whole region, but with different densities/pressures in the three regions,
the other is adopting three different materials in the three regions.
For the case adopting multiple materials,
the remeshing operations on the discontinuity interfaces have to obey the remapping rules and criterions on
multi-material interfaces, as described in Figs. \ref{fig.SwapMapOnInterface}-\ref{fig.MergeMapOnInterface} and in Sec. \ref{sec.RemeshAlgorithm}.
While for the case adopting single material, the remeshing operations act without those constraints.
A comparative study of those two cases can assess the performance of the remeshing algorithm on multi-material interfaces.

\begin{figure}[htbp]
\centering
  \includegraphics[width=0.9\textwidth]{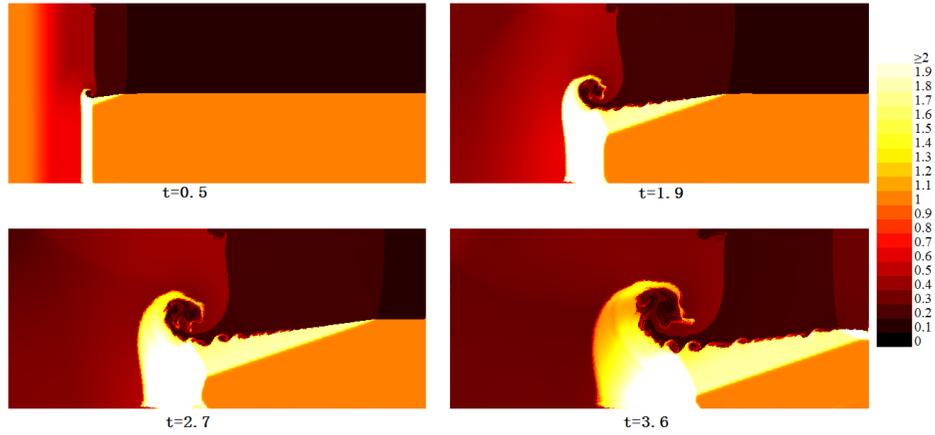}\\
\caption{Density maps of triple-point evolution from the simulation adopting the single-material approach and the compensation matter flow.}\label{fig.TripleSingleMatterWithFlow}
\end{figure}

Fig. \ref{fig.TripleSingleMatterWithFlow} shows density maps of the triple-point evolution from the simulation adopting the single-material approach and with the compensation matter flow.
Due to the discrepancy in the $\Omega_B$ and $\Omega_C$, the shockwaves propagate with different speeds in the two domains.
This creates a shear along the initial discontinuity and a vortex rolls up.
Fine structures of the vortex are presented by the pictures in Fig. \ref{fig.TripleSingleMatterWithFlow}.

\begin{figure}[htbp]
\centering
  \includegraphics[width=0.6\textwidth]{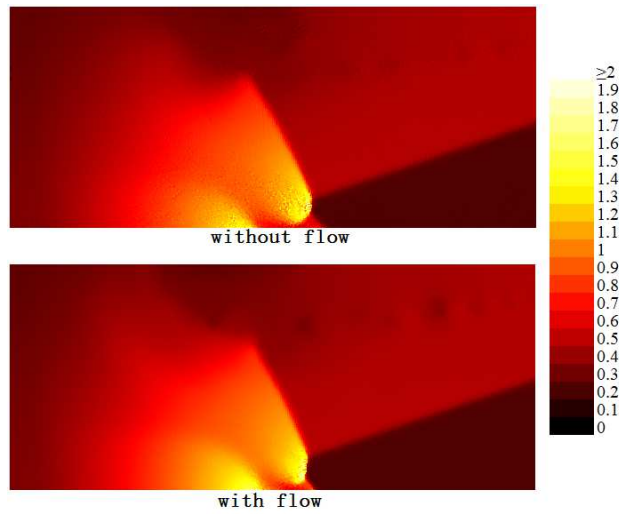}
\caption{Comparison of the pressure maps of triple-point simulations without and with the compensation matter flow, at $t=3.60$.}\label{fig.TriplePressureFlowCompare}
\end{figure}

Fig. \ref{fig.TriplePressureFlowCompare} compares the simulation adopting single-material approach but without the compensation mass flow
with the simulation in Fig. \ref{fig.TripleSingleMatterWithFlow}.
For the simulation without the compensation, there are visible oscillations in the pressure map,
while the pressure distribution is much smoother for simulation adopting the compensation matter flow.

\begin{figure}[htbp]
\centering
  \includegraphics[width=0.6\textwidth]{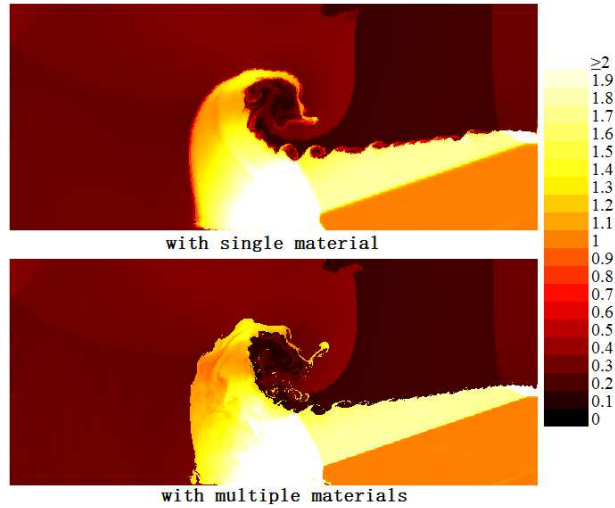}
\caption{Comparison of the pressure maps of triple-point simulations with single material and multiple materials, at $t=3.60$.}\label{fig.TriplePressureMultiMaterial}
\end{figure}

Fig. \ref{fig.TriplePressureMultiMaterial} compares the simulation with the compensation mass flow but adopting the multiple-materials approach 
with the simulation in Fig. \ref{fig.TripleSingleMatterWithFlow}.
The two simulations are alike in general.
Some major differences of them are observed at the interfaces.
For the simulation adopting multiple materials, the Kelvin-Helmholtz instability at the interface is not so easy to roll up.
The reason is considered to be that:
since the compensation matter flow which represents the effects of edge-bending is not allowed to occur between different materials at the interfaces,
the fluids motion at the interfaces is not so well captured in the multi-material case.

\subsection{Heavy-light fluids convection}

At the initial state,
a heavy fluid is placed upon a light fluid,
and a perturbation is set on the initial shape of their interface.
The initial densities of the heavy and light fluids are $1.0$ and $0.1$.
Both fluids are barotropic fluid.
The EOS is of the form $p=k(\rho-\rho_0)/\rho_0$,
with $k=1000.0$ for both the heavy and light fluids.
The surface tension forces of interface between the two fluids and the free surfaces of the fluids are all set
to be 0.2. 

\begin{figure}[htpb]
\centering
  \includegraphics[width=0.3\textwidth]{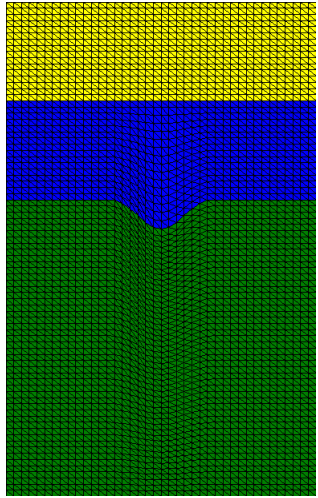}\\
  \caption{Mesh adopted for the heavy-light fluids convection simulation.}\label{fig.DoubleFluidsMesh}
\end{figure}

A triangular mesh with $40\times 80\times 2$ elements is constructed for the simulation, as shown in
Fig. \ref{fig.DoubleFluidsMesh}.
The top layer is a vacuum background (which is replaced by a fluid with very a small density),
the middle layer is the heavy fluid,
and the bottom layer is the light fluid.

\begin{figure}[htbp]
\centering
  \includegraphics[width=0.9\textwidth]{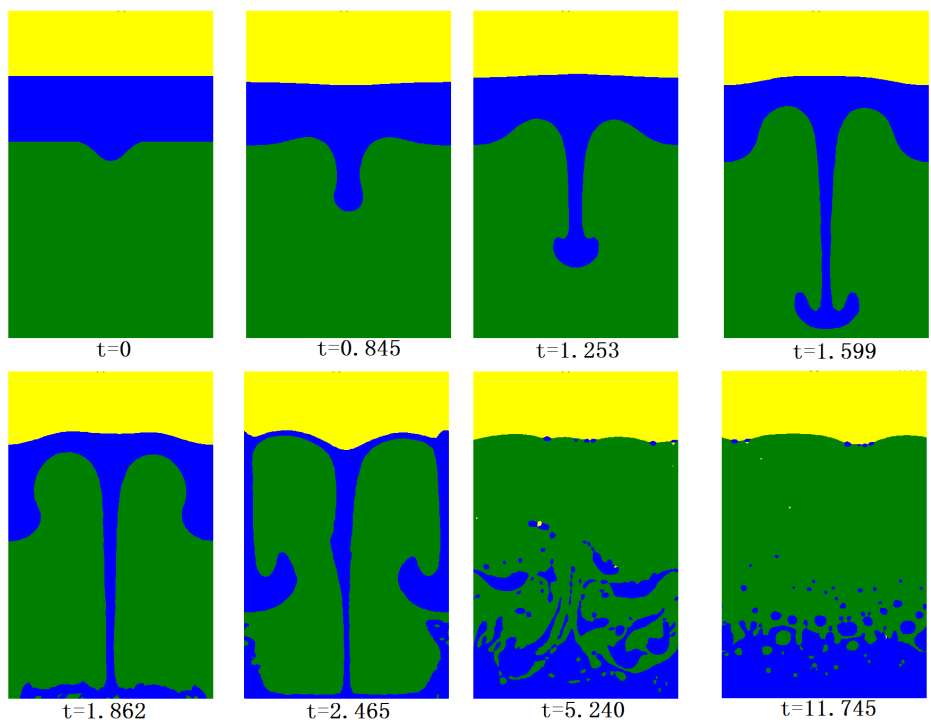}
\caption{Pictures of the heavy-light fluids convection evolution.}\label{fig.DoubleFluidsConvection}
\end{figure}

Pictures of the evolution process are shown in Fig. \ref{fig.DoubleFluidsConvection}.
In the early stage of the evolution, as the heavy fluid dips into the light fluid,
the Rayleigh-Taylor instability gradually grows into the form of a mushroom.
After the heavy fluid touches the bottom, the light fluid is pushed apart and lifted up.
In the end, the light fluid floats over the heavy one, and a steady state is reached.
These stages are full of extreme deforming and topology variable phenomenons such as the bubbles forming and breakage,
whose simulation is usually considered to be a hard task for the classic Lagrangian methods.

\section{Conclusion}

In this paper, two major improvements are made for the SGH Lagrangian scheme.
The first is the construction of a dynamic local remeshing method for preventing mesh distortion in SGH Lagrangian simulation.
The approach for handling multi-material interface grids in the remeshing scheme is found to work fine
by comparing the single-material and multiple-materials approaches in the triple-point simulations, and also from the successful 
simulation of the extremely deforming multi-material process of a heavy fluid dipping into a light fluid.
The second improvement is the introduction of a compensation matter flow derived from edge-bending for mitigating
the checkerboard oscillation problem in Lagrangian simulations.
The effectiveness of this compensation method is proved by the simulations of the sod, Noh, and triple-point problems.

However, there also exist some unsatisfying phenomenons.
Why does the density distribution has larger oscillation after the introduction of 
the compensation matter flow for the simulation of the Noh test problem?
How to extend the compensation matter flow method to the multi-material interfaces
so as to better capture the fluids behaviors around multi-material interfaces?
These problems remain to be studied in future works.

{\em Acknowledgements:} This work was supported partly by the Foundation of China Academy of Engineering Physics (No. 2015B0201023) 
and the National Natural Science Foundation of China (Nos. 11405167, 11571293 and 11672276).

%% References


\bibliographystyle{elsarticle-num}
\bibliography{<your-bib-database>}



\end{document}